\documentclass[nofootinbib,aps,prc,reprint,showpacs,superscriptaddress,twocolumn,notitlepage]{revtex4-1}

\usepackage{graphicx,color} 
\usepackage{bm}       
\usepackage{bbm}
\usepackage{amsmath} 
\usepackage[dvipsnames]{xcolor}
\usepackage{ulem}
\usepackage{multirow}
\usepackage{float} 
\newcommand {\nc} {\newcommand}

\nc{\ninej}[9]{\left\{\begin{array}{ccc} #1 & #2 & #3 \\ #4 & #5 & #6 \\ #7 & #8 & #9 \\ \end{array}\right\}}
\nc{\sixj}[6]{\left\{\begin{array}{ccc} #1 & #2 & #3 \\ #4 & #5 & #6 \\ \end{array}\right\}}
\nc{\threej}[6]{ \left( \begin{array}{ccc} #1 & #2 & #3 \\ #4 & #5 & #6 \\ \end{array} \right) }
\nc{\half}{\frac{1}{2}}

\nc{\lla}{\left\langle}
\nc{\rra}{\right\rangle}

\nc {\IR} [1]{\textcolor{red}{#1}}
\nc {\IB} [1]{\textcolor{blue}{#1}}         
\nc {\IP} [1]{\textcolor{magenta}{#1}}
\nc {\IM} [1]{\textcolor{Bittersweet}{#1}} 
\nc {\IE} [1]{\textcolor{Plum}{#1}}        
\nc {\IG} [1]{\textcolor{OliveGreen}{#1}}

\begin{document}

\title{Separable character of \textit{ab initio} No-Core Shell Model one-body densities}

\author{J.~Foy}
\affiliation{Institute of Nuclear and Particle Physics, and Department of Physics and Astronomy, Ohio University, Athens, OH 45701, USA}

\author{Ch.~Elster}
\affiliation{Institute of Nuclear and Particle Physics, and Department of Physics and Astronomy, Ohio University, Athens, OH 45701, USA}

\author{P.~Maris}
\affiliation{Department of Physics and Astronomy, Iowa State University, Ames, IA 50011, USA}

\author{S.P. Weppner}
\affiliation{Natural Sciences, Eckerd College, St. Petersburg, FL 33711, USA }

\author{S.K. Bogner}
\affiliation{Facility for Rare Isotope Beams and Department of Physics and Astronomy, Michigan State University, East Lansing, Michigan 48824, USA}
\affiliation{Institute of Nuclear and Particle Physics, and Department of Physics and Astronomy, Ohio University, Athens, OH 45701, USA}

\date{\today}

\begin{abstract}
Motivated by recent findings on the separability of optical potentials that are derived from folding off-shell densities with off-shell nucleon-nucleon amplitudes, we study the off-shell character of one-body density matrices created within the No-Core Shell Model (NCSM). 
Concentrating on nuclei with a $0^+$ ground state from $^4$He through $^{48}$Ca, we investigate the off-shell character of their one-body density matrices in momentum space when using the momentum transfer and the average momentum as variables. 
A singular value decomposition of the one-body density matrices reveals that they can be characterized by only very few terms, depending on the mass number of the nucleus. 
These findings are independent of the nucleon-nucleon interactions employed, as well as of computational specifics, such as grid spacing and size of the model space. 

\end{abstract}


\maketitle

\section{Introduction and Motivation}
\label{sec:intro}

Over the last decade, developments of the nucleon-nucleon ($NN$) and three-nucleon ($3N$) interactions derived from chiral effective field theory have yielded major progress~\cite{Epelbaum:2005pn,Machleidt:2011zz,Epelbaum:2019kcf,Piarulli:2019cqu,Machleidt:2024bwl}.
They, together with the use of massively parallel computing resources, have placed large-scale {\it ab initio} simulations at the frontier of nuclear structure and reaction explorations, see, e.g. \cite{Elhatisari:2022zrb,Piarulli:2022ulk,King:2024zbv,LENPIC:2018ewt}.
Among other successful many-body theories, the {\it ab initio} No-Core Shell Model (NCSM) or No-Core Configuration Interaction (NCCI) approach~\cite{Barrett:2013nh}, has over the last decade taken center stage in the development of microscopic tools for studying the structure of atomic nuclei up to $A\simeq 20$.
Following the developments in {\it ab initio} structure theory, rigorous calculations of effective folding nucleon-nucleus ($NA$) interactions for elastic scattering of protons or neutrons from nuclei in the same mass region were developed~\cite{Burrows:2018ggt,Burrows:2020qvu,Baker:2021izp,Baker:2021iqy,Gennari:2017yez,Vorabbi:2021kho,Vorabbi:2020cgf,Arellano:2022tsi,Baker:2024wtn} based on the leading-order in the spectator expansion of multiple scattering theory~\cite{Siciliano:1977zz,Ernst:1977gb,Tandy:1980zz}.

Recently, Arellano and Blanchon~\cite{Arellano:2024xoc,Arellano:2022tsi} found that folding optical potentials for elastic nucleon-nucleus ($NA$) scattering from $0^+$ nuclei for target mass numbers in the range of $40 \leq A \leq 208$ exhibit a separable character in their radial and off-shell (nonlocal) features, which proved to be universal in this mass regime for beam energies between 40 and 400~MeV. This is indeed a striking outcome of their studies. Since folding optical potentials are computed by integrating over off-shell density matrices and off-shell nucleon-nucleon amplitudes, we want to concentrate in this work on one of the ingredients of this folding integral, namely the off-shell nuclear one-body density.  If this quantity turns out to be separable, then the findings of Arellano and Blanchon can be more easily understood, since the result of an integral over a separable function and any arbitrary function is separable. 

In this work we concentrate on translationally invariant off-shell one-body densities obtained from NCSM calculations for nuclei with $0^+$ ground states~\cite{Burrows:2017wqn} in the mass range $4\leq A \leq 48$ obtained with the NNLO$_{\rm opt}$ chiral interaction~\cite{Ekstrom13} and the Daejeon16~\cite{Shirokov:2016ead} potential, as well as a chiral $NN$ plus $3N$ LENPIC interaction~\cite{Reinert:2017usi}. To determine if those off-shell density matrices exhibit a separable structure, we employ the technique of singular value decomposition (SVD), which will give information about the rank of separability. The SVD is a powerful mathematical tool with applications across many areas of science and engineering~\cite{BruntonKutz,Strang2019}. It enables dimensional reduction for expensive high-fidelity models, low-rank approximations of large matrices and tensors, and the identification of dominant correlations in large-scale data sets. In nuclear physics, SVD techniques have played a central role in the development of fast and accurate emulators for scattering and bound-state calculations~\cite{Tichai:2022mqn,Maldonado:2025ftg, Davisonthesis}, the construction of low-rank approximations of nucleon-nucleon ($NN$) and 3-nucleon ($3N$) interactions~\cite{Tichai:2021rtv, Zhu:2021pis}, and dimensionality reduction for {\it ab initio} many-body calculations~\cite{Papenbrock:2003bj,Frosini:2024ajq}.     

The construction of translationally invariant off-shell density matrices in momentum space is briefly reviewed in Sec.~\ref{nonlocald} for the convenience of the reader. In addition, specific results for the closed-shell nuclei $^4$He, $^{16}$O, and $^{40}$Ca are shown.  In Sec.~\ref {svd} we apply the SVD to off-shell density matrices for closed as well as open-shell nuclei and study the systematics of the singular values as a function of the nuclear mass up to $^{48}$Ca.  In Sec.~\ref{reconstruction} we show that a separable representation with a given rank indeed captures the properties of the off-shell density matrices, and in Sec.~\ref{eigenvectors} we study in detail properties of the singular vectors arising in the SVD. For Sec.~\ref{extreme}, we construct an `extreme shell-model' to extrapolate our insights to nuclei heavier than $^{48}$Ca.  We conclude in Sec.~\ref{conclusions}.


\section{Off-shell Momentum Space NCSM densities for 0$^+$ Ground States}
\label{nonlocald}

The concept of an off-shell (nonlocal), translationally invariant one-body density in momentum space derived from NCSM calculations has been extensively discussed in a previous publication~\cite{Burrows:2017wqn}. 
Here, we do not intend to repeat the formulation in all its details.
However, for the reader's convenience, we summarize the essential steps we undertake to obtain the $0^+$ ground state densities we study.
In the NCSM, the many-body wave function $\Psi$ is expanded in a basis of Slater determinants of single-particle states, which uses space-fixed ($sf$) single-particle coordinates.  
In terms of an $A$-body wavefunction $\Psi_{JM}({\bf p}_1, {\bf p}_2, \ldots, {\bf p}_A)$, for a state with total angular momentum $J$ and magnetic projection $M$, the space-fixed off-shell one-body density matrix (OBDM) in momentum space is given by
\begin{eqnarray}
\lefteqn{  {\bar{\rho}}^{JM}_{sf} ({\mathbf p}',{\mathbf p})  =  \int \Psi^*_{JM}({\mathbf p}', {\mathbf p}_2, \ldots, {\mathbf p}_A) }
\nonumber \\
    && \times \; \Psi_{JM}({\mathbf p}, {\mathbf p}_2, \ldots, {\mathbf p}_A) \; d^3 {\mathbf p}_2 \ldots d^3 {\mathbf p}_A  \,,
\end{eqnarray}
where ${\bf p}$ and ${\bf p}'$ are the initial and final momenta of a single nucleon.

Note that the NCSM wave functions, and thus the space-fixed OBDM, include a center-of-mass (c.m.) contribution that needs to be removed.  
In practice, we use harmonic oscillator (HO) single-particle states, with radial quantum number $n$, orbital quantum number $l$, angular momentum $j$, magnetic projection $m$, and HO energy $\hbar\omega$ (or equivalently, HO length $b=\sqrt{\frac{\hbar^2}{m_N \, \hbar\omega}}$, with $m_N$ being the nucleon mass), in combination with a truncation of the many-body basis space based on the total number of HO quanta, characterized by $N_{\rm max}$, the maximum number of oscillator quanta above the valence shell for the given nucleus.  
With this truncation, the NCSM wavefunctions factorize into a relative wavefunction and an HO c.m. wavefunction, which allows us to remove the c.m. contribution from the space-fixed OBDM.

The space-fixed OBDM, ${\bar{\rho}}^{JM}_{sf} ({\bf p}',{\bf p})$, can most easily be obtained from the NCSM wavefunctions by first performing a multipole expansion
\begin{eqnarray}
\lefteqn{  {\bar{\rho}}^{JM}_{sf} (\mathbf{p},\mathbf{p}') = }
\nonumber \\
  && \sum_{K=0}^{2J}(-1)^{J-M}\begin{pmatrix} J & K & J \\ -M & 0 & M \end{pmatrix}  \; \rho^{J(K)}_{sf} (\mathbf{p},\mathbf{p}') \,, 
\end{eqnarray}
with the reduced one-body densities, $\rho^{J(K)}_{sf} ({\bf p},{\bf p}')$, given in terms of the reduced one-body density matrix elements of the NCSM wavefunctions, $\langle A \lambda J||(a_{n'l'j'}^{\textsuperscript{\textdagger}}\tilde{a}_{nlj})^{\left(K\right)}|| A\lambda J \rangle$, where $a$ and $a^{\textsuperscript{\textdagger}}$ are single-nucleon annihilation and creation operators, and $\lambda$ stands for any additional quantum numbers of the NCSM eigenstate of interest (see Ref.~\cite{Cockrell:2012vd} for more details).  The general expression for the reduced one-body densities is given by
\begin{eqnarray}
\lefteqn{\rho^{J(K)}_{sf} (\mathbf{p},\mathbf{p}') = \sum_{nln'l'jj'}(-1)^{l+l'+j+\frac{1}{2}} }
\nonumber \\
    && \times \; R_{n,l}(p)R_{n',l'}(p') \; \mathcal{Y}^{* l' l}_{K0}\left(\hat{p},\hat{p}'\right) \sqrt{(2j+1)(2j'+1)} 
\nonumber \\    
    && \times \; \begin{Bmatrix} l' & l & K \\ j & j' & \frac{1}{2}\end{Bmatrix}
    \left\langle A \lambda J\left|\left|\left(a_{n'l'j'}^{\textsuperscript{\textdagger}}\tilde{a}_{nlj}\right)^{\left(K\right)}\right|\right| A\lambda J \right\rangle ,
\label{eq:ROBDMppprime}
\end{eqnarray}
where $R_{n,l}(p)$  and $R_{n',l'}(p')$ are the radial HO wavefunctions, and $\mathcal{Y}^{* l' l}_{K0}\left(\hat{p},\hat{p}'\right)$ is the bipolar harmonic of $\hat{p}$ and $\hat{p}'$ (see Ref.~\cite{Burrows:2017wqn} for more details).

To remove the c.m. contribution from $\rho^{J(K)}_{sf} (\mathbf{p},\mathbf{p}')$ and obtain the translationally-invariant off-shell one-body density, we need to make a coordinate transformation to the momentum transfer, ${\bf q} = {\bf p}' - {\bf p}$, and the average momentum, $\bm{\mathcal{K}}=\frac{1}{2} (\mathbf{p}+\mathbf{p}')$.
Using the Talmi--Moshinski transformations, we can express (see Appendix B of Ref.~\cite{Burrows:2017wqn})
\begin{eqnarray}
\lefteqn{ R_{n,l}(p)R_{n',l'}(p')\mathcal{Y}^{* l' l}_{K0}\left(\hat{p},\hat{p}'\right) =} \nonumber \\
&&\sum_{n_{q}n_{\mathcal{K}}l_{q}l_{\mathcal{K}}}\left\langle n_{q}l_{q},n_{\mathcal{K}}l_{\mathcal{K}}:K|n'l',nl:K\right\rangle_{d=1}
\nonumber \\
&& \times \; R_{n_{q},l_{q}}(q)R_{n_{\mathcal{K}},l_{\mathcal{K}}}(\mathcal{K})\mathcal{Y}^{* l_{q} l_{\mathcal{K}}}_{K0}\left(\hat{q},\hat{\mathcal{K}}\right) \,.
\label{eq:TM}
\end{eqnarray}
Substituting Eq.~(\ref{eq:TM}) into Eq.~(\ref{eq:ROBDMppprime}) gives the reduced space-fixed off-shell OBDMs
as functions of the momentum transfer ${\bf q}$ and average momentum ${\bm {\mathcal K}}$,
\begin{eqnarray}
 \lefteqn{ \rho_{sf}^{J(K)}\left({\bf q},\bm{\mathcal{K}}\right) =
 \sum_{l_q l_{\mathcal{K}}}  \mathcal{Y}_{K0}^{*l_{q}l_{\mathcal{K}}}(\hat{q},\hat{\mathcal{K}})  }
 \nonumber \\
&& \sum_{n_q,n_{\mathcal{K}},nln'l'jj'}
 \left(-1\right)^{j+\frac{1}{2}}\left\langle n_{q}l_{q},n_{\mathcal{K}}l_{\mathcal{K}}:K|n'l',nl:K\right\rangle_{d=1} \nonumber \\
&&  \times \; R_{n_{q}l_{q}}\left(q\right)R_{n_{\mathcal{K}}l_{\mathcal{K}}}\left(\mathcal{K}\right) \; \sqrt{\left(2j+1\right)\left(2j'+1\right)} \nonumber \\
&&  \times \; \begin{Bmatrix} l' & l & K \\ j & j' & \frac{1}{2} \end{Bmatrix}\left\langle A \lambda J\left|\left|\left(a_{n'l'j'}^{\textsuperscript{\textdagger}}\tilde{a}_{nlj}\right)^{\left(K\right)}\right|\right|\
 A\lambda J \right\rangle \, .
\label{eq:ROBDMqKappa}
\end{eqnarray}

In NCSM calculations in a HO basis with $N_{\rm max}$ truncation, the wave function in single-particle coordinates exactly factorizes into a translationally-invariant wave function and a c.m. wave function; furthermore, this c.m. contribution is a $1s$ HO wavefunction,
\begin{equation}
|\Psi JM\rangle_{sf} = |\Psi JM\rangle \otimes |\phi_{\rm c.m.} \; 1s\rangle \,.
\label{eq:cmfactorization}
\end{equation}
Following Ref.~\cite{Burrows:2017wqn}, the c.m. contribution to $\rho^{J(K)}_{sf} (\mathbf{p},\mathbf{p}')$  can be calculated exactly, leading to a simple analytic function of $q^2$, so that the reduced translationally-invariant off-shell density is given by
\begin{eqnarray}
  \rho^{J(K)}(\mathbf{q},\bm{\mathcal{K}}) &=& e^{\frac{q^2\,b^2}{4 \, A}}
  \; \rho^{J(K)}_{sf}(\mathbf{q},\bm{\mathcal{K}}),
\label{eq:ROBDMti}
\end{eqnarray}
with $b^2 = \frac{\hbar^2}{m_N \,\hbar\omega}$.

In the following, we will concentrate on nuclei with $0^+$ ground states, which only have a monopole contribution.  For simplicity, we will omit the multipole index $K$ and the index $J$ from the reduced translational invariant off-shell density from here on.  Furthermore, except for $^{48}$Ca, we only consider $N=Z$ nuclei; we therefore only show neutron densities (the proton densities are nearly identical to the neutron densities for all of the $N=Z$ nuclei that we consider here).

\begin{figure}[tb]
    \centering
\includegraphics[width=0.8\columnwidth]{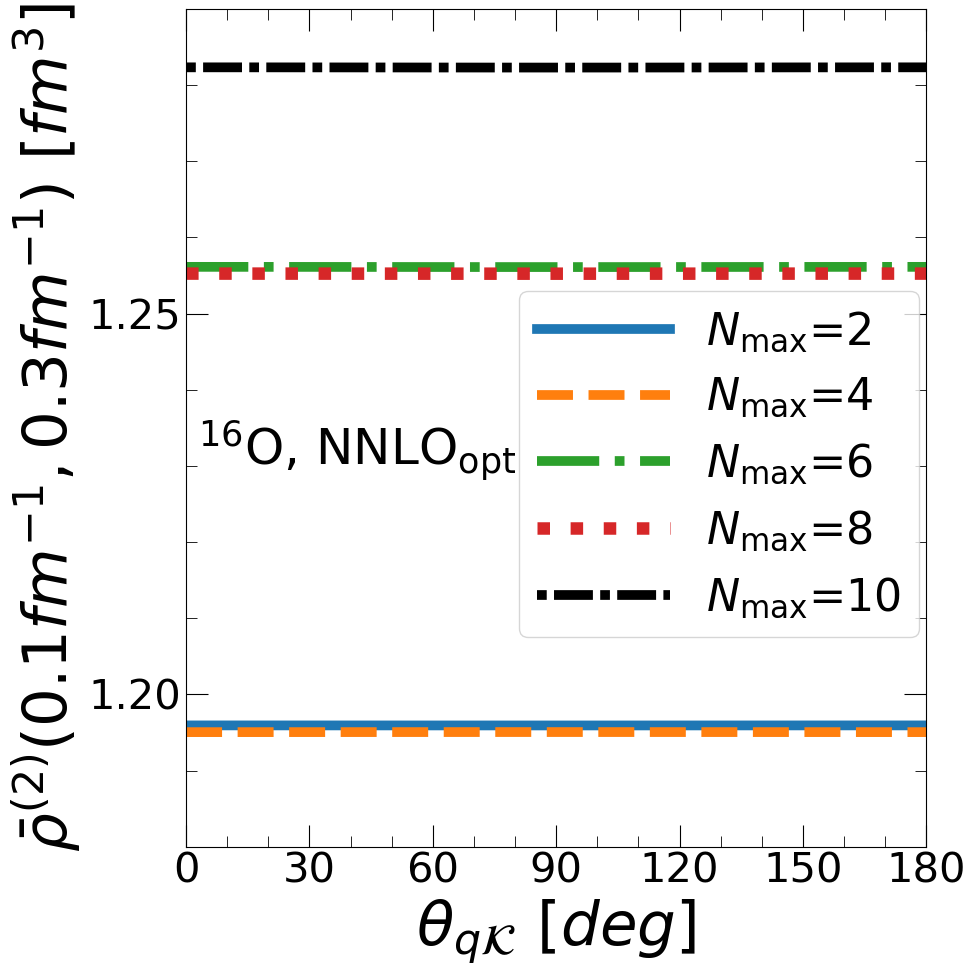} 
\caption{The neutron density of $^{16}$O calculated with the NNLO$_{\rm opt}$ chiral potential~\cite{Ekstrom13} as function of the angle $\theta_{q{\mathcal K}}$ at fixed values of $q$ and $\mathcal K$ for a series of $N_{\rm max}$ values.  All calculations use $\hbar\omega = 20$~MeV.}
\label{fig2-1}
\end{figure}
The reduced off-shell OBDM $\rho ({\bf q}, {\bm {\mathcal K}})$ is a function of three variables; namely, the magnitudes $q = |{\bf q}|$ and ${\mathcal K} = |\bm {\mathcal K}|$, as well as the angle $\theta_{q{\mathcal K}}$ between the two vectors.   
When fixing specific values of the magnitudes $q$ and ${\mathcal K}$, we observe that $\rho ({\bf q}, {\bm {\mathcal K}})$ does not depend on the angle between the two vectors, independent of the value of $N_{\rm max}$,  independent of the value of $\hbar\omega$, and independent of the interaction.  In Fig.~\ref{fig2-1} we show this independence of the angle $\theta_{q{\mathcal K}}$ for the neutron density of $^{16}$O obtained from the NNLO$_{\rm opt}$ chiral potential~\cite{Ekstrom13} as an example (the corresponding proton density looks essentially the same).
We found qualitatively the same behavior for all $0^+$ nuclei we studied, as well as for different $NN$ interactions we employed.  Adding a $3N$ interaction does not change this behavior either.
This finding strongly suggests that the monopole term of the off-shell momentum-space OBDM is indeed separable in the variables $q$ and ${\mathcal K}$.
However, the angle independence does not give any information about the rank of separability.

\begin{figure}[tb]
    \centering
\includegraphics[width=0.99\columnwidth]{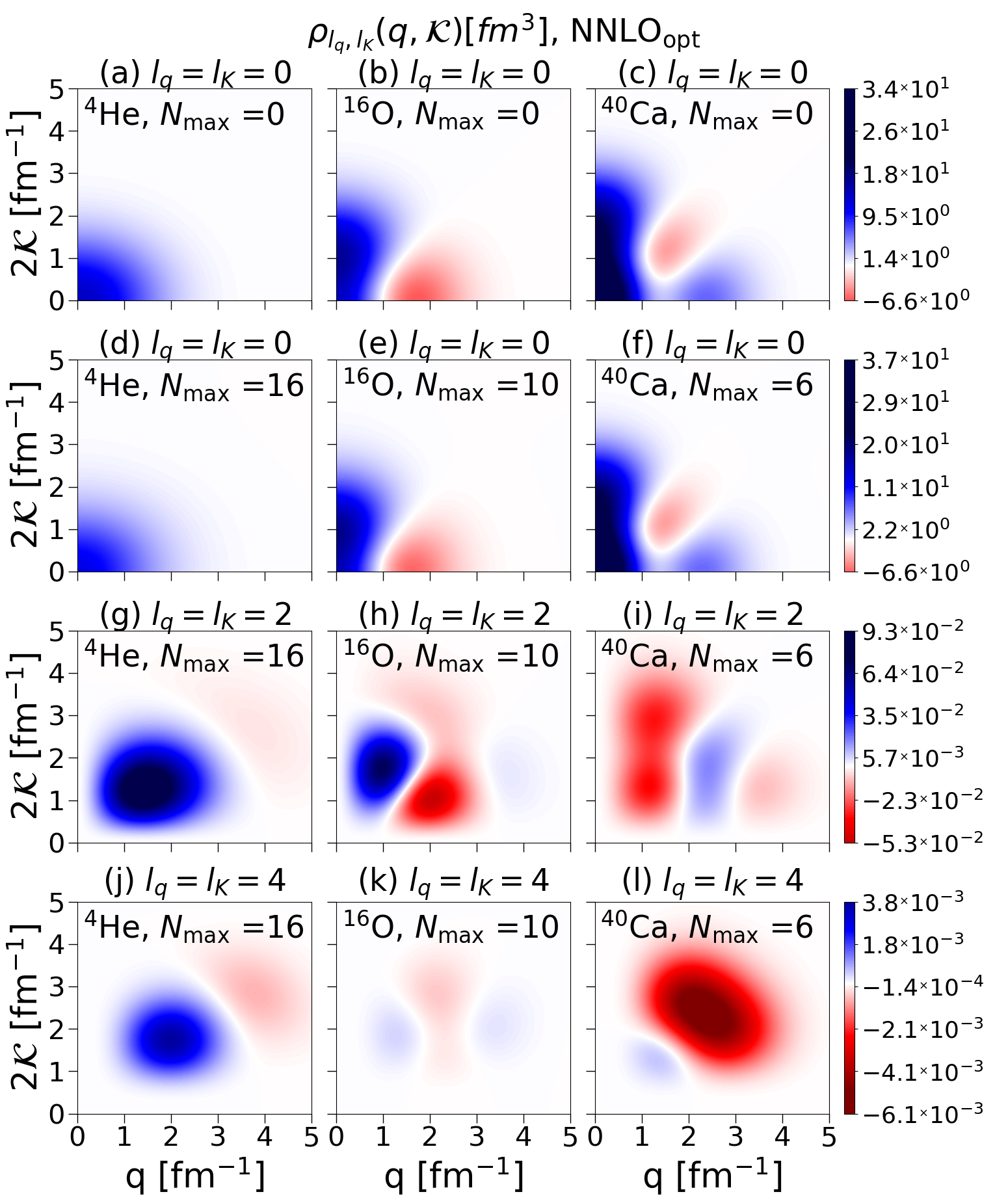}
\caption{The $K=0$ component of the translationally-invariant off-shell one-body density obtained from NCSM calculations based on the NNLO$_{\rm opt}$ chiral potential~\cite{Ekstrom13} for the neutron distributions of $^4$He (left column), $^{16}$O (middle column), and $^{40}$Ca (right column) as function of the momenta $q$ and ${\mathcal K}$. Each row shows $l_q=l_{\mathcal K}$
components from 0 to 4.
All calculations use $\hbar\omega = 20$~MeV. The values of $N_{\rm max}$ for the different calculations are given in each panel. 
Note the change of scales for the higher partial waves. 
}
\label{fig2-2}
\end{figure}
The off-shell density in Eq.~(\ref{eq:ROBDMqKappa}) is written in a form that shows the explicit sum over angular momenta $l_q$ and $l_{\mathcal K}$. 
For the monopole term, $K=0$, the constraints from Eq.~(\ref{eq:TM}) require $l_q = l_{\mathcal K}$.  
We also point out that the contributions of odd values of $l_q$ cancel out exactly due to the Talmi--Moshinsky brackets~\cite{Kamuntavicius:2001pf,moshinsky:1996:harmonic}. 
In addition, the Wigner-6j symbol forces $j=j'$ and $l=l'$, which leads to a straightforward partial wave representation of the off-shell density,
\begin{eqnarray}
 \lefteqn{\rho_{l_{q} l_{\mathcal K}} \left(q,\mathcal{K}\right) = \;
 e^{\frac{q^{2}b^{2}}{4A}} \sum_{n_{q}n_{\mathcal K} nln'j} (-1)^{j+\frac{1}{2}} \left(2j+1\right)} \nonumber \\
&&  
\times \left\langle n_{q}l_{q},n_{\mathcal{K}}l_{q}:0|n'l,nl:0\right\rangle_{d=1} 
R_{n_{q}l_{q}}\left(q\right)R_{n_{\mathcal{K}}l_{q}}\left(\mathcal{K}\right)
\nonumber \\ 
&& 
\times \begin{Bmatrix} l & l & 0 \\ j & j & \frac{1}{2} \end{Bmatrix}\left\langle A \lambda J\left|\left|\left(a_{n'lj}^{\textsuperscript{\textdagger}}\tilde{a}_{nlj}\right)^{\left(0\right)}\right|\right| A\lambda J \right\rangle  \,.
\label{eq:ROBDMpartwavedec}
\end{eqnarray}
In Fig.~\ref{fig2-2} we present the partial wave decomposition for three closed-shell nuclei, $^4$He, $^{16}$O, and $^{40}$Ca, from $l_q=l_{\mathcal K} = 0$ to $4$. The calculations are based on the NNLO$_{\rm opt}$ chiral potential~\cite{Ekstrom13}.  
The corresponding $N_{\rm max}$ for each nucleus is given in the panels.  We chose the same color scale for each angular momentum.
In the top row, we depict the results from calculations at $N_{\rm max}=0$, which means that we only have contributions from $l_q=l_{\mathcal K}=0$, since there are no contributions from higher angular momenta.

The following three rows show the angular momentum contributions for calculations with larger values of $N_{\rm max}$, as shown in the different panels.  
First, we observe that the $l_q=l_{\mathcal K}=0$ contribution dominates for all three nuclei, and the general shape is visually the same as for the $N_{\rm max}=0$ calculation, while the scale is slightly different.  
The next higher angular momentum contribution, $l_q=l_{\mathcal K}=2$, is already two orders of magnitude smaller than the $l_q=l_{\mathcal K}=0$ contribution, indicating that the lowest angular momentum contributions dominate the off-shell density.  
This trend continues with the next-higher angular momentum, $l_q=l_{\mathcal K}=4$, and is independent of the interaction.
 
Finally, we want to draw attention to some characteristic features of the momentum space one-body density when the momentum transfer $q$ goes to zero.  In that case, the vectors ${\mathbf p}$ and ${\mathbf p}'$ become the same, and ${\bm {\mathcal K}}$ becomes equal to ${\bf p}$.  
In that limit, the function $\rho(0,\bm{\mathcal K})$ corresponds to the probability density of finding a nucleon with momentum ${\mathbf p}={\bm {\mathcal K}}$ inside the nucleus.  
Note that this is analogous to the local density in coordinate space, which can be obtained from the nonlocal density $\tilde\rho(\bm{\zeta}, \mathbf{Z})$ in the limit when $\mathbf{Z}$ goes to zero, and which corresponds to the probability density of finding a nucleon at position $\bm{\zeta}$ relative to the center-of-mass of the nucleus.
Here we use the variables $\bm{\zeta}=\frac{1}{2} ({\bf r}'+{\bf r})$ and $\mathbf{Z}={\bf r}'-{\bf r}$ following the notation of Ref.~\cite{Burrows:2017wqn}, and $\tilde\rho(\bm{\zeta}, \mathbf{Z})$ is the Fourier transform of $\rho({\bf q},\bm{\mathcal K})$.
Indeed, in Fig.~\ref{fig2-2} one can see that $\rho(0,{\bm{\mathcal K}})$ for all shown nuclei is positive definite as it should be for a probability distribution.  
We also note that only $l_q=l_{\mathcal K}=0$ can contribute to the slice at $q=0$ as a result of the $q^l$ behavior for small $q$ of partial wave amplitudes.



\section{Singular Value Decomposition of the off-shell one-body density}
\label{svd}

As shown in Fig.~\ref{fig2-1}, the off-shell density for nuclei with $0^+$ ground states is independent of the angle between the two vectors ${\bf q}$ and ${\bm {\mathcal K}}$, which points to it being separable in those two momenta. 
Let us assume it is separable; then it can be written in the form 
\begin{equation}
\rho(q,\mathcal{K}) \approx \sum_{r}^{R} \sigma_{r} \, f_{r}(q) \, g_{r}(\mathcal{K}) \, . 
\label{eq:3.1}
\end{equation}
Here, $\sigma_{r}$ are scaling constants, and $R$ is the rank of separability.
To determine this rank $R$, we employ the method of singular value decomposition (SVD).
The SVD (see e.g. Ref.~\cite{Strang} or Sec. IV of Ref.~\cite{Maldonado:2025ftg}) is a matrix decomposition of any real (or complex) matrix into two orthogonal (or unitary) matrices, $\mathbf{u}$ and $\mathbf{v}^{T}$, and a diagonal matrix with non-negative entries in decreasing order, the singular values $\sigma_r$.
 
For the numerical decomposition, we start from the partial wave decomposed one-body density matrix and use the $l_{q}=l_{\mathcal{K}}=0$ partial wave, since we already know that the $s$-wave gives the dominant contribution.  
(We confirmed that the numerical results are identical when using the $l_{q}=l_{\mathcal{K}}=0$ partial wave or the density at specific values of the angle $\theta_{q{\mathcal K}}$ between $\bf q$ and $\bm{\mathcal K}$.)
Next, we discretize the function $\rho_{l_{q},l_{\mathcal{K}}=0}(q,\mathcal{K})$ on an $N \times N$ grid of momenta $q$ and $\mathcal{K}$, and define a square matrix $\mathbf{M}$ as
\begin{equation}
    M_{i,j} := \rho_{l_{q},l_{\mathcal{K}}=0}(q_{j},\mathcal{K}_{i}),
\label{eq:3.2}
\end{equation} 
where the indices $i$ and $j$ run from 1 to $N$. 
For simplicity of notation, the subscripts $l_{q},l_{\mathcal{K}}=0$ will be omitted from the following. 
The complete decomposition is thus of the form 
\begin{equation}
    M_{i,j} = \sum_{r=1}^{N}\sigma_{r} \, \mathbf{u}_{r}(\mathcal{K}_{i}) \, \mathbf{v}^{T}_{r}(q_{j}) \,,
\label{eq:3.3}
\end{equation} 
where $\sigma_{r}$ is the r-th singular value, $\mathbf{u}_{r}(\mathcal{K}_{i})$ is $i$th element of the $r$th left singular vector and $\mathbf{v}^{T}_{r}(q_{j})$ is the $j$th element of the $r$th right singular vector.

The orthonormal column vectors $\mathbf{u}_r$ and $\mathbf{v}_r$ are the left and right singular vectors, respectively. 
The span of $\{ {\bf u}_i \}$ associated with $\sigma_i$ equals to the span of the matrix $\mathbf{M}$ of Eq.~(\ref{eq:3.3}).
Truncating the SVD to a specific value $R \leq N$ then gives the best rank-$R$ approximation of the matrix $\mathbf{M}$ in the Frobenius norm.
A standard option to determine how many singular values and corresponding left and right vectors are needed to represent the original matrix up to a given error is using a preserved variance (PV), defined as
\begin{equation}
\textnormal{PV} (R) \equiv \frac{\sum_{r=1}^{R} \sigma_{r}^{2}}{\sum_{r=1}^N\sigma_{r}^{2}} \, , 
\label{eq:3.4}
\end{equation}
from which we can define the tolerance $\eta$ as
\begin{equation}
\eta({R}) = 1 - \textnormal{PV} (R) \, .
\label{eq:3.4b}
\end{equation}
Here $R\leq N$ refers to the desired rank-$R$ approximation, and $\eta(R)$ is a (small) truncation tolerance. 
We set our threshold for this tolerance to $\eta(R)=10^{-5}$, a value that is sufficiently small for a reasonably accurate representation of the complete off-shell one-body density. At the same time, it is still well above the numerical noise in our calculations.\footnote{Note that the NCSM calculations of the input OBDMEs in Eq.~(\ref{eq:ROBDMpartwavedec}), which provide the input for the off-shell one-body densities, are carried out in single precision.}
All calculations presented in this work are carried out with a grid size $N=301$, and a grid spacing of $0.0\Bar{3}$~fm$^{-1}$, corresponding to a cutoff on $q$ and $\mathcal{K}$ of 10~fm$^{-1}$.  We verified that our results for $\eta(R)$ are stable as long as we have a grid spacing of 0.1~fm$^{-1}$ per dimension and the momentum cutoff is at least 10.0~fm$^{-1}$.
Thus, for a given $R$, if the value of the tolerance, $\eta(R)$, falls below the threshold,  we consider this value of $R$ as the minimum rank necessary for an accurate separable ansatz of the complete off-shell one-body density for a given nucleus. On the other hand, if $\eta(R) > 10^{-5}$, the rank $R$ needs to be increased. We choose this tolerance threshold based on the ability to reproduce the RMS radius to three figures and on the qualitative match between the associated density figures. 

\begin{figure}[tb]
    \centering
\includegraphics[width=0.8\columnwidth]{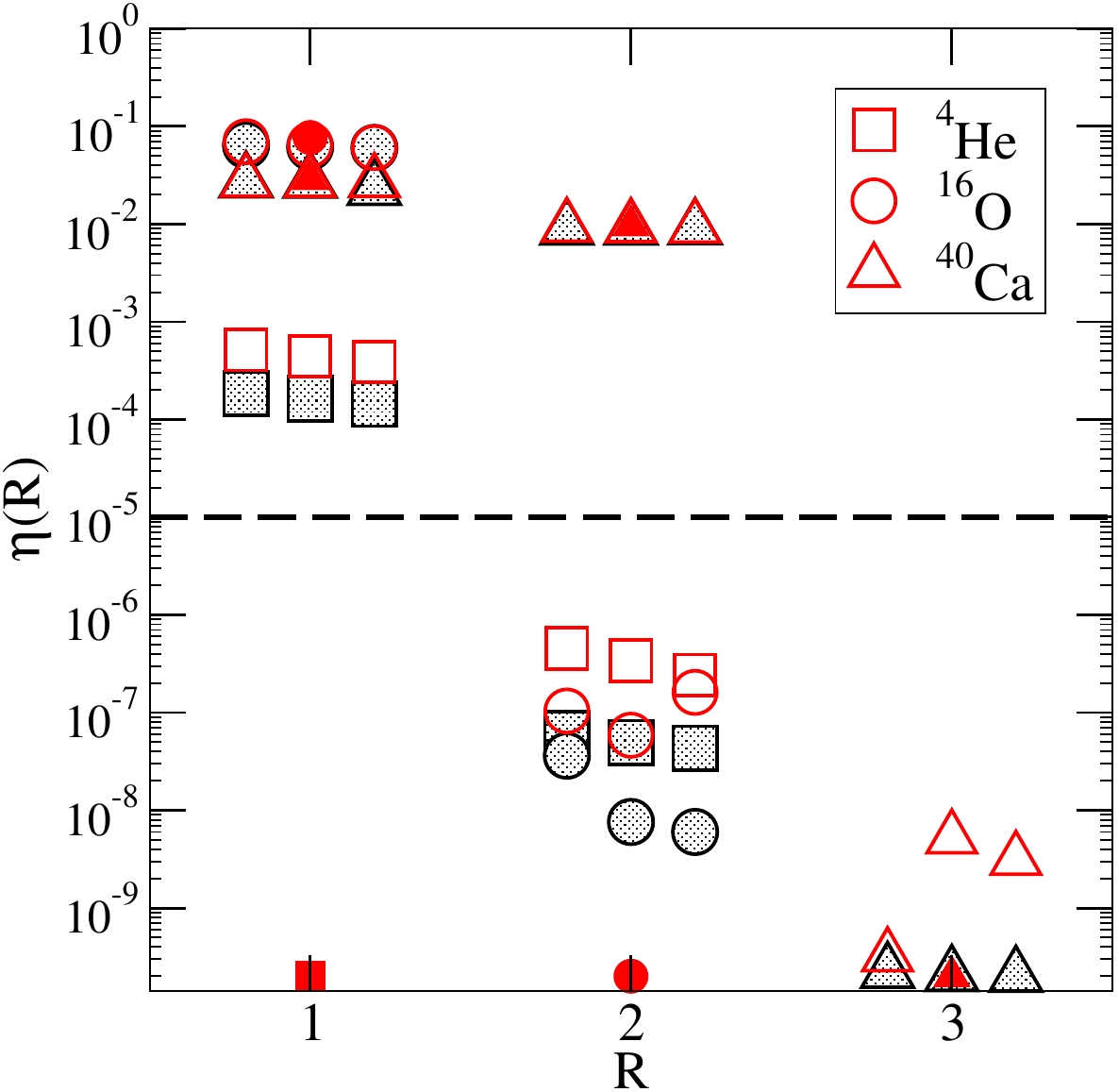}
\caption{The calculated values of $\eta (R)$ as function of $R$ for $^4$He 
($N_{\rm{max}}=16$), $^{16}$O ($N_{\rm {max}}=10$), and $^{40}$Ca ($N_{\rm{max}}=6$) 
based on the NNLO$_{\rm opt}$ chiral potential~\cite{Ekstrom13}  (red open symbols) and the Daejeon16 potential~\cite{Shirokov:2016ead} (grey hashed symbols). 
The filled symbols indicate calculations using $N_{\rm max}=0$ and $\hbar \omega=20$~MeV. 
A detailed explanation of the symbols is given in the text.
The dashed line indicates our tolerance threshold of $10^{-5}$ for $\eta (R)$.
}
\label{svd-1}
\end{figure}

As a first step, we consider the closed shell nuclei $^4$He, $^{16}$O, and $^{40}$Ca, where successively the $s$-, $p$-, and $sd$-shells are filled.  
The resulting values of $\eta (R)$ are shown in Fig.~\ref{svd-1} as a function of $R$ for two different interactions, the NNLO$_{\rm opt}$ chiral interaction~\cite{Ekstrom13} and the Daejeon16 (DJ16)~\cite{Shirokov:2016ead} potential.
Let us first consider the calculations using $N_{\rm max}=0$, which are independent of the interaction and carried out for $\hbar\omega=20$~MeV. 
In the case of $^4$He, $R=1$ is already well below the required tolerance threshold, indicating that the off-shell density of $^4$He is a rank-1 separable function at $N_{\rm max}=0$. 
For $^{16}$O, one needs to sum up two SVD values to fall below the tolerance threshold, making the off-shell density of $^{16}$O at $N_{\rm max}=0$ a rank-2 separable function. 
In the case of $^{40}$Ca at $N_{\rm max}=0$, one needs to go to $R=3$ before $\eta(R)$ drops below our threshold, making the off-shell density a rank-3 separable function.
This suggests that the minimal rank necessary to represent the off-shell density may be related to the underlying shell structure.

For more realistic cases, we turn to calculations with higher values of $N_{\rm max}$ for each of the closed shell nuclei.
The values of $\eta(R)$ for calculations performed with an oscillator energy of $\hbar \omega=20$~MeV are centered directly on the integer value of $R$.
We also carried out calculations with $\hbar \omega=16$ and $24$~MeV for NNLO$_{\rm opt}$, and with $\hbar\omega=17.5$ and $22.5$~MeV for Daejeon16. 
To make those points better visible, they are shifted horizontally (left for $\hbar \omega=16$ and $17.5$~MeV and right for $\hbar \omega=22.5$ and $24$~MeV).

First, we notice that the off-shell density of $^4$He computed with $N_{\max} > 0$, although producing a small value of $\eta(R)$, needs a second SVD value to fall below the tolerance threshold.
Though the figure depicts $N_{\rm max}=16$, this feature arises immediately when increasing $N_{\rm max}$ from zero to two.
Therefore, we conclude that converged calculations of the off-shell density of $^4$He need to be represented by rank-2 separable functions and cannot be accurately represented by a rank-1 function as one may naively assume based on the $N_{\rm max} = 0$ result, and when thinking of $^4$He as a closed $s$-shell nucleus.  This is most likely related to the fact that $^4$He is not actually a pure $s$-shell nucleus, but contains contributions from higher shells as well.

For $^{16}$O, we see in Fig.~\ref{svd-1} that $\eta(R)$ is well above our tolerance threshold of $10^{-5}$ at $R=1$, but for $R=2$ it drops well below this threshold, independent of the HO parameter, and independent of the $NN$ interaction employed.  This indicates that the off-shell density for $^{16}$O can be represented by a rank-2 separable function.  
Notice that this is the case both at $N_{\max} = 0$ and at $N_{\max} = 10$.
Thus, we can conclude that the off-shell density for $^{16}$O is a rank-2 separable function, independent of the HO parameter, and independent of the $NN$ interaction employed.  We also know that the nucleus $^{16}$O can be envisioned as having a filled $s$- and $p$-shell.  This corroborates our intuition 
that the rank of the one-body density may be related to the shell structure.

Finally, in the nucleus $^{40}$Ca, the $sd$-shell is filled in addition to $s$- and $p$-shells.
And indeed, Fig.~\ref{svd-1} shows that $\eta (R=3)$ falls below the tolerance threshold, indicating that a rank-3 separable function should be sufficient to approximate the off-shell density of $^{40}$Ca, both at $N_{\max} = 0$ and at $N_{\max} = 6$.
We will explicitly investigate these statements in more detail in the following Sections.

\begin{figure}[tb]
\centering
\includegraphics[width=0.97\columnwidth]{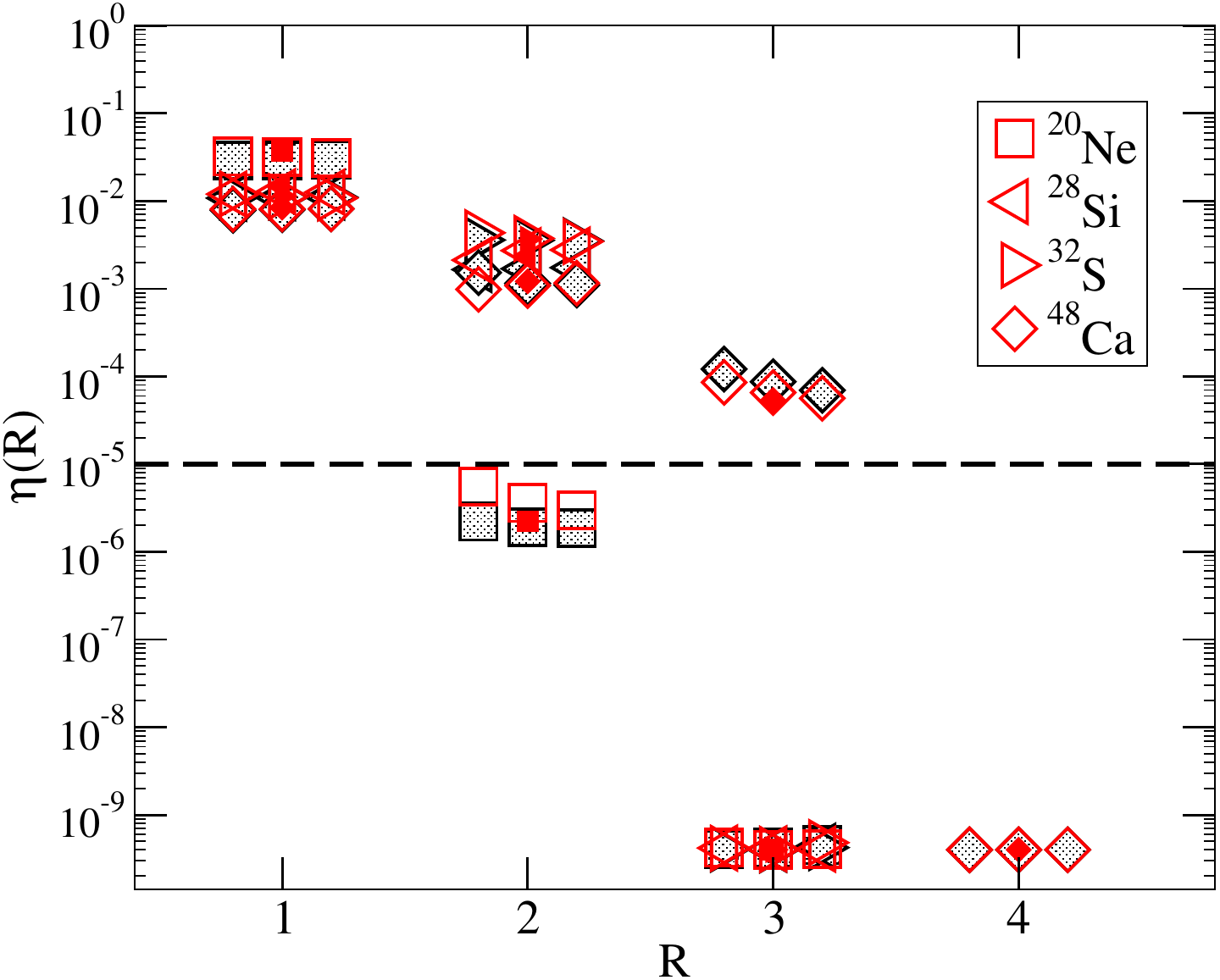}
\caption{The calculated values of $\eta(R)$ as a function of $R$ for $^{20}$Ne,
$^{28}$Si, $^{32}$S (all at $N_{\max}=4$), and $^{48}$Ca (at $N_{\max}=2$) based on the NNLO$_{\rm opt}$ (red open symbols) and the Daejeon16 (grey hashed symbols) potentials. The filled symbols indicate calculations using $N_{\rm max}$=0 at $\hbar\omega = 20$~MeV with NNLO$_{\rm opt}$. 
The dashed line indicates our tolerance threshold of $10^{-5}$.
}
\label{svd-2}
\end{figure}
Next, we focus our interest on open-shell nuclei in the mass regime from $A=20$ to $A=48$, and show in Fig.~\ref{svd-2} our results for $\eta(R)$ for the $0^+$ ground states of $^{20}$Ne, $^{28}$Si, $^{32}$S, and $^{48}$Ca.
The open symbols represent values of $\eta(R)$ obtained with the NNLO$_{\rm opt}$ potential calculated with $N_{\rm max}=4$ for the $sd$-shell nuclei $^{20}$Ne, $^{28}$Si, and $^{32}$S, and with $N_{\rm max}=2$ for $^{48}$Ca; the hashed symbols correspond to results with the Daejeon16 potential, at the same $N_{\max}$.
Note that the corresponding results for the $sd$-shell nuclei $^{20}$Ne, $^{28}$Si, $^{32}$ at $N_{\max}=2$ would be indistinguishable from the results shown at $N_{\max}=4$ on this logarithmic scale, which spans 10 orders of magnitude.
For completeness, we also include the NNLO$_{\rm opt}$ potential results using $N_{\rm max}=0$ at $\hbar\omega = 20$~MeV, and show those with filled symbols.
As in Fig.~\ref{svd-1}, our results for $\eta(R)$ obtained with an oscillator energy of $\hbar\omega = 20$~MeV are centered directly on the integer value of $R$.
In addition, Fig.~\ref{svd-2} contains calculations with different values of the oscillator energy, $\hbar\omega$=16 and 24~MeV for NNLO$_{\rm opt}$ and with $\hbar\omega$=17.5 and 22.5~MeV for Daejeon16.
To make those points more visible, they are shifted horizontally (left for $\hbar\omega$=16 and 17.5~MeV and right for $\hbar\omega$=22.5 and 24~MeV).

We realize that this figure looks rather cluttered; however, this is part of the point we want to make: qualitatively, our results with different truncation parameters $N_{\max}$, different HO parameters $\hbar\omega$, and different interactions all agree.
For $^{20}$Ne, with only a few nucleons in the $sd$-shell, the value of $\eta(R=2)$ is slightly below the chosen tolerance threshold, suggesting that the off-shell density of this nucleus can still be described reasonably well by a rank-2 separable approximation.  
We will study this in more detail in the next Section. 
For the other open $sd$-shell nuclei, $^{28}$Si and $^{32}$Ne, one needs $R=3$ for $\eta(R)$ to drop below the tolerance threshold of $10^{-5}$ independent of oscillator length, the basis cutoff $N_{\rm max}$, and the interaction.
This means that for $sd$-shell nuclei, the off-shell one-body density can be represented by a rank-3 separable function in the momentum transfer $q$ and the average momentum $\mathcal K$. 
Only if one moves beyond the filled $sd$ shells of 20 neutrons and 20 protons in $^{40}$Ca and into the $pf$ shell, does one need to go to $R=4$  for $\eta(R)$ to drop below the tolerance threshold, as can be seen from our results for $^{48}$Ca, with the $f_{7/2}$ sub-shell filled with eight neutrons.

Up to now, we have only shown calculations for $J=0$ ground states, obtained with $NN$ interactions -- and qualitatively, our results appear to be independent of the specific $NN$ potential.  
This raises two questions, namely, whether or not a $3N$ interaction might change this picture, and whether or not the same conclusion holds for states with nonzero $J$.
To explore the effects of $3N$ interactions we used the LENPIC Semilocal Momentum Space (SMS) chiral EFT interaction at N$^3$LO, with a chiral EFT cutoff of $\Lambda = 500$~MeV \cite{Reinert:2017usi}, softened using the Similarity Renormalization Group (SRG) to $\alpha = 0.08$~fm$^4$, which gives a good description of $p$-shell nuclei~\cite{LENPIC:2022cyu,Maris:2023esu}.
\begin{figure}[tb]
\centering
\includegraphics[width=0.97\columnwidth]{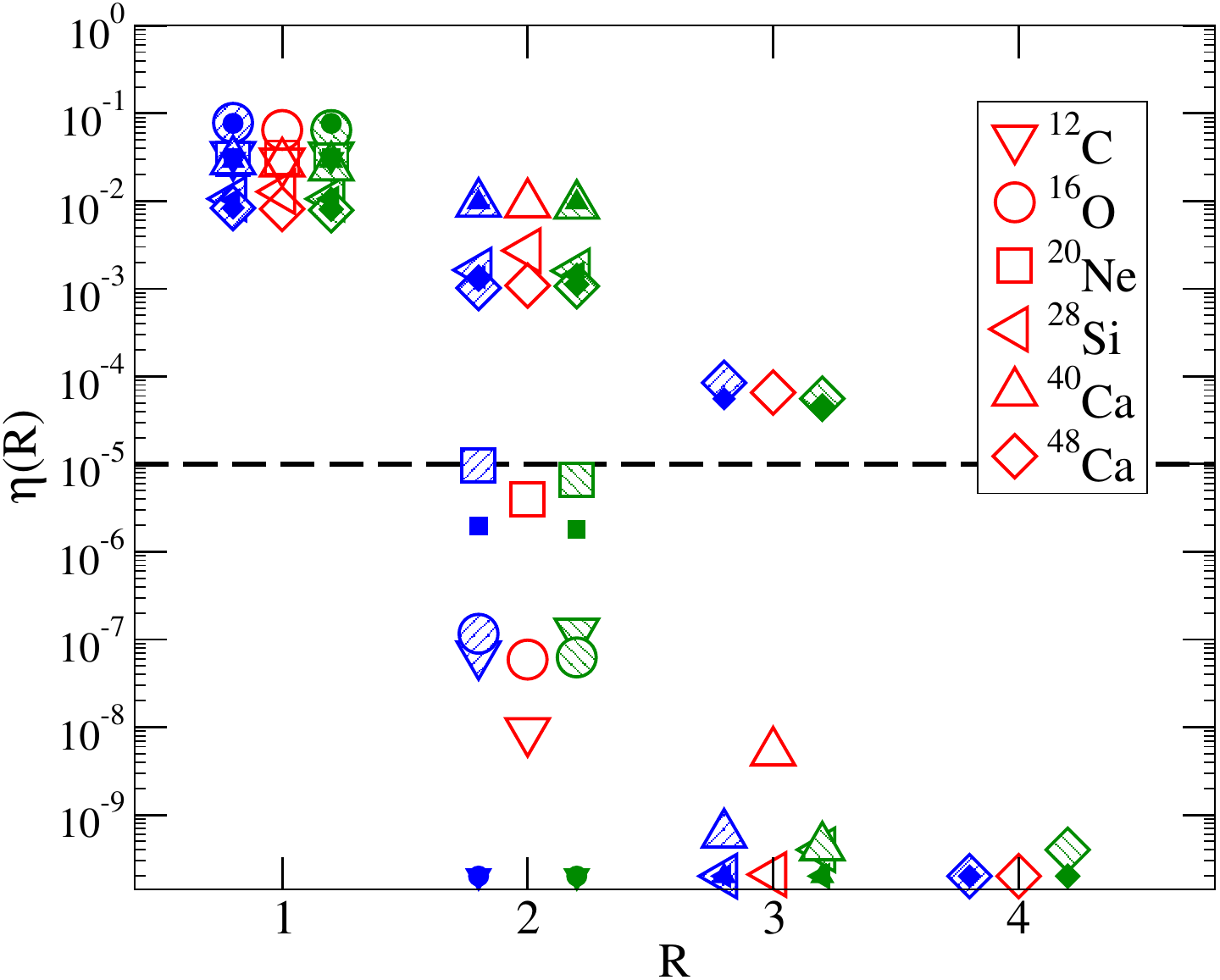}
\caption{The calculated value of $\eta(R)$ as function of $R$ for $^{12}$C, $^{16}$O, $^{20}$Ne (all at $N_{\rm max}=6$), $^{28}$Si ($N_{\rm max}=2$), $^{40}$Ca ($N_{\rm max}=4$), and $^{48}$Ca ($N_{\rm max}=2$).
The blue symbols, shifted to the left, represent the LENPIC SMS chiral N$^3$LO $NN$ potential, while the green symbols, shifted to the right, represent calculations including the corresponding $3N$ interactions~\cite{Reinert:2017usi}.  Symbols filled with a solid color represent $N_{\rm max} = 0$.
For comparison, the open symbols represent the NNLO$_{\rm opt}$ interaction using $N_{\rm max}=10$ for $^{12}$C and $^{16}$O, $N_{\rm max}=6$ for $^{20}$Ne and $^{40}$Ca, $N_{\rm max}=4$ for $^{28}$Si, and $N_{\rm max}=2$ for $^{48}$Ca.
The dashed line indicates the tolerance threshold of $\eta(R)=10^{-5}$, and all calculations were performed at $\hbar\omega=20$~MeV.
}
\label{svd-3}
\end{figure}
In Fig.~\ref{svd-3} we show our results for the ground states of $^{12}$C, $^{16}$O, $^{20}$Ne, $^{28}$Si, $^{40}$Ca, and $^{48}$Ca with the LENPIC SMS interaction using only the $NN$ interactions (symbols shifted to the left) as well as including the corresponding $3N$ interactions (shifted to the right).  As a reference, the calculations with the NNLO$_{\rm opt}$ potential are shown centered on $R$.  From the discussion of the results presented in Figs.~\ref{svd-1} and~\ref{svd-2}, we conclude that the same considerations for the ranks that are needed to capture the behavior of the off-shell densities hold when adding $3N$ interactions. 

Finally, in Appendix~\ref{appendixA} we show that the one-body densities of the $J=3$ ground state of $^{10}$B and the $J=0$ ground state of $^{12}$C are both rank-2 separable.  Furthermore, we demonstrate in that appendix that our results are reasonably well converged with respect to N$_{\rm max}$ and $\hbar\omega$ as long as $\eta(R)$ is above our threshold of $10^{-5}$.  On the other hand, once $\eta(R)$ drops below this threshold, it becomes more sensitive to our many-body basis truncation parameter $N_{\max}$ and the HO scale $\hbar\omega$. 

These observations let us conclude that our main results appear to be universal: the off-shell one-body density is separable in $q$ and $\mathcal K$, with the rank increasing with the number filled shells, from rank-2 in the $p$-shell to rank-3 in the $sd$-shell to rank-4 in the $pf$-shell.  This behavior is independent of the numerical details such as the HO parameter $\hbar\omega$ or the many-body truncation parameter $N_{\max}$, and qualitatively, our results do
not depend on the specific $NN$ interaction employed, nor do they change when $3N$ interactions are included.

\section{Separable representation}
\label{reconstruction}

In the previous Section, we demonstrated that the singular value decomposition of the NCSM off-shell one-body density for nuclei with a 0$^+$ ground state suggests a classification based on the rank of separability, which is related to their structure, or rather, to the number of occupied shells.
According to this classification, the off-shell one-body density for nuclei with $\max(N,Z) \leq 8$ can be represented by rank-2 separable functions, and nuclei with higher masses, up to $\max(N,Z) \leq 20$ by rank-3 separable functions, independent of the interaction and independent of computational details such as the oscillator parameter $\hbar\omega$ or the NCSM truncation parameter.
\begin{figure*}[bt]
    \begin{flushleft}
    \includegraphics[width=0.70\textwidth]{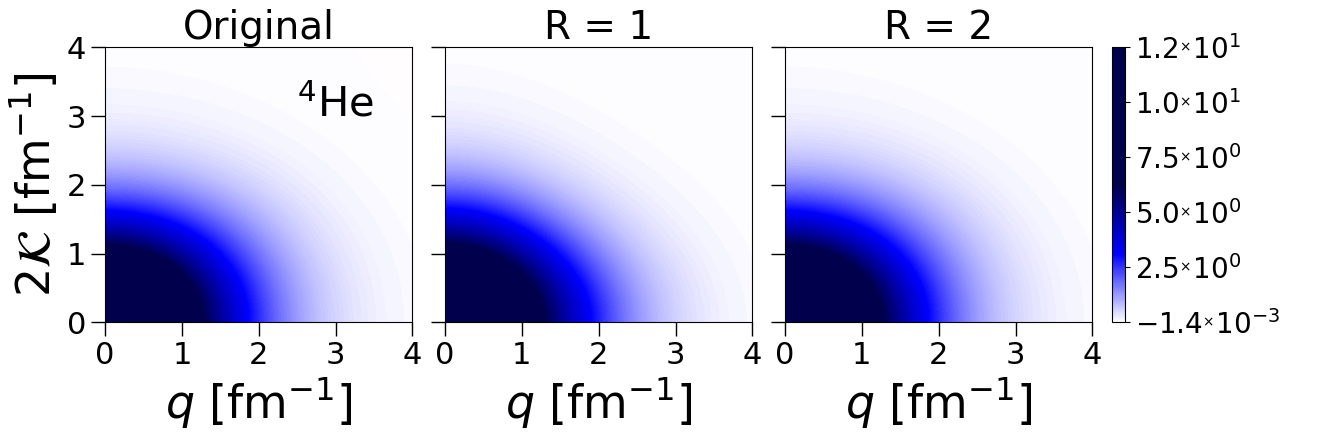} \\
  \includegraphics[width=0.70\textwidth]{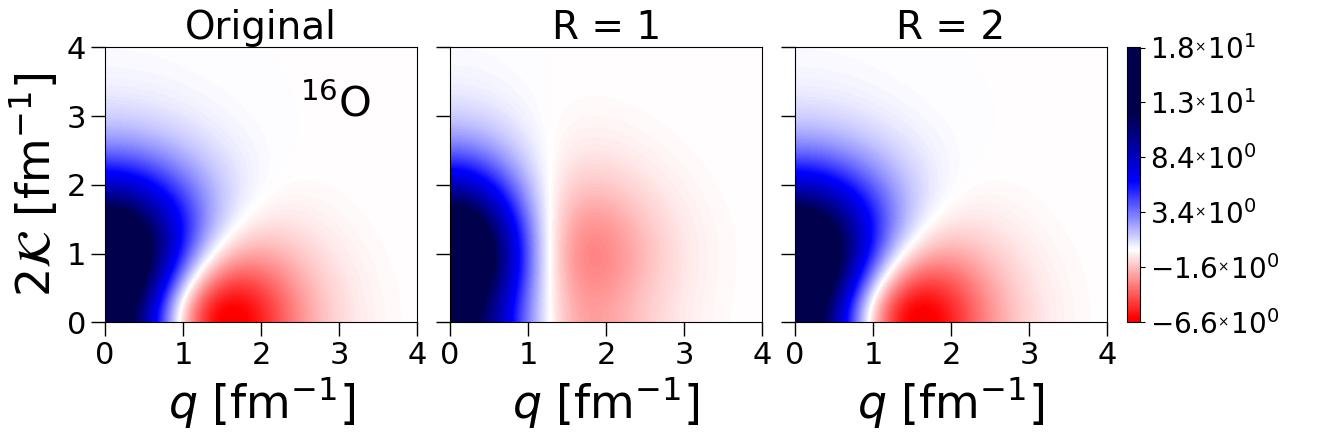} \\
  \includegraphics[width=0.90\textwidth]{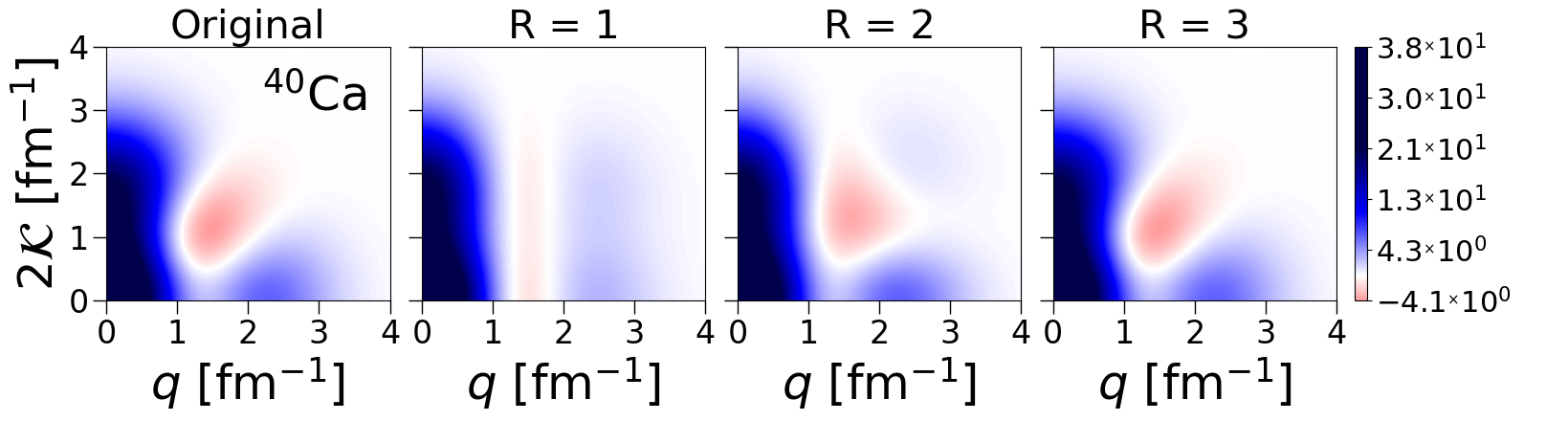}
    \end{flushleft}
    \caption{The $l_{q}=l_{\mathcal{K}}=0$ partial waves of the off-shell one-body densities $\rho(q,\mathcal{K})$ for $^{4}$He (top row), $^{16}$O (middle row), and $^{40}$Ca (bottom row). 
    The left panels, labeled 'Original', depict $\rho_{l_{q}=l_{\mathcal{K}}=0}(q,\mathcal{K})$ for each nucleus;
    the subsequent panels show rank-$R$ separable approximations, starting from $R=1$ up to the minimal rank for which $\eta(R) < 10^{-5}$ for each nucleus.
    All calculations are carried out with NNLO$_{\rm opt}$ at $\hbar\omega = 20$~MeV,
    using the same values of $N_{\rm max}$ as given in the caption of Fig.~\ref{svd-1}.
    }   
\label{fig4-1}
\end{figure*}

In this Section, we present numerical checks that this is indeed the case.
First, we begin with a qualitative visual inspection of how the off-shell one-body density is constructed by adding separable terms. 
And again, we concentrate on the $l_{q}=l_{\mathcal{K}}=0$ partial wave as being the dominant one and define
\begin{equation}
 \rho_{l_{q}=l_{\mathcal{K}}=0}(q,\mathcal{K}) \approx \rho_R(q,\mathcal{K}) \equiv \sum_{r=1}^R \sigma_{r}\mathbf{u}_{r}(\mathcal{K})\mathbf{v}^T_{r}(q).
\label{eq:}
\end{equation}
By changing the upper limit $R$ in the sum over the number of ranks, we can see how the separable approximation to the off-shell density changes as a function of $R$.

For this demonstration we concentrate on the closed-shell nuclei $^4$He, $^{16}$O, and $^{40}$Ca, using the NNLO$_{\rm opt}$ chiral interaction. 
The values of $\eta(R)$ for those nuclei (together with the corresponding $N_{\rm max}$) have already been shown in Fig.~\ref{svd-1}. 
Figure~\ref{fig4-1} shows the off-shell one-body densities $\rho_{l_{q}=l_{\mathcal{K}}=0}(q,\mathcal{K})$ in the left-most column (labeled 'Original'), followed by the rank-1 and rank-2 SVD approximations; and for $^{40}$Ca, we also show the rank-3 SVD approximation.

As already pointed out in the previous Section, the nucleus $^4$He is somewhat special, namely for $N_{\rm max}=0$ the value of $\eta(R)$ drops below the tolerance already for $R=1$, while for $N_{\rm max}=16$ (which is what is shown in Fig.~\ref{fig4-1}) we need $R=2$ for $\eta(R)$ to fall below our tolerance.
However, from the qualitative pictures in the top row of Fig.~\ref{fig4-1}, the necessity of a rank-2 approximation is not apparent for $^4$He.
This is quite different for $^{16}$O and $^{40}$Ca, where a rank-1 approximation does not resemble the structure of the actual off-shell one-body density. 
According to the values of $\eta(R)$ from Fig.~\ref{svd-1}, the off-shell density of $^{16}$O can be approximated by a rank-2 separable function, while for $^{40}$Ca, one needs a rank-3 function.  
This is indeed qualitatively confirmed by the two-dimensional density plots in the second and third rows of Fig.~\ref{fig4-1}.

\begin{figure}[tb]
\centering
\includegraphics[width=0.99\columnwidth]{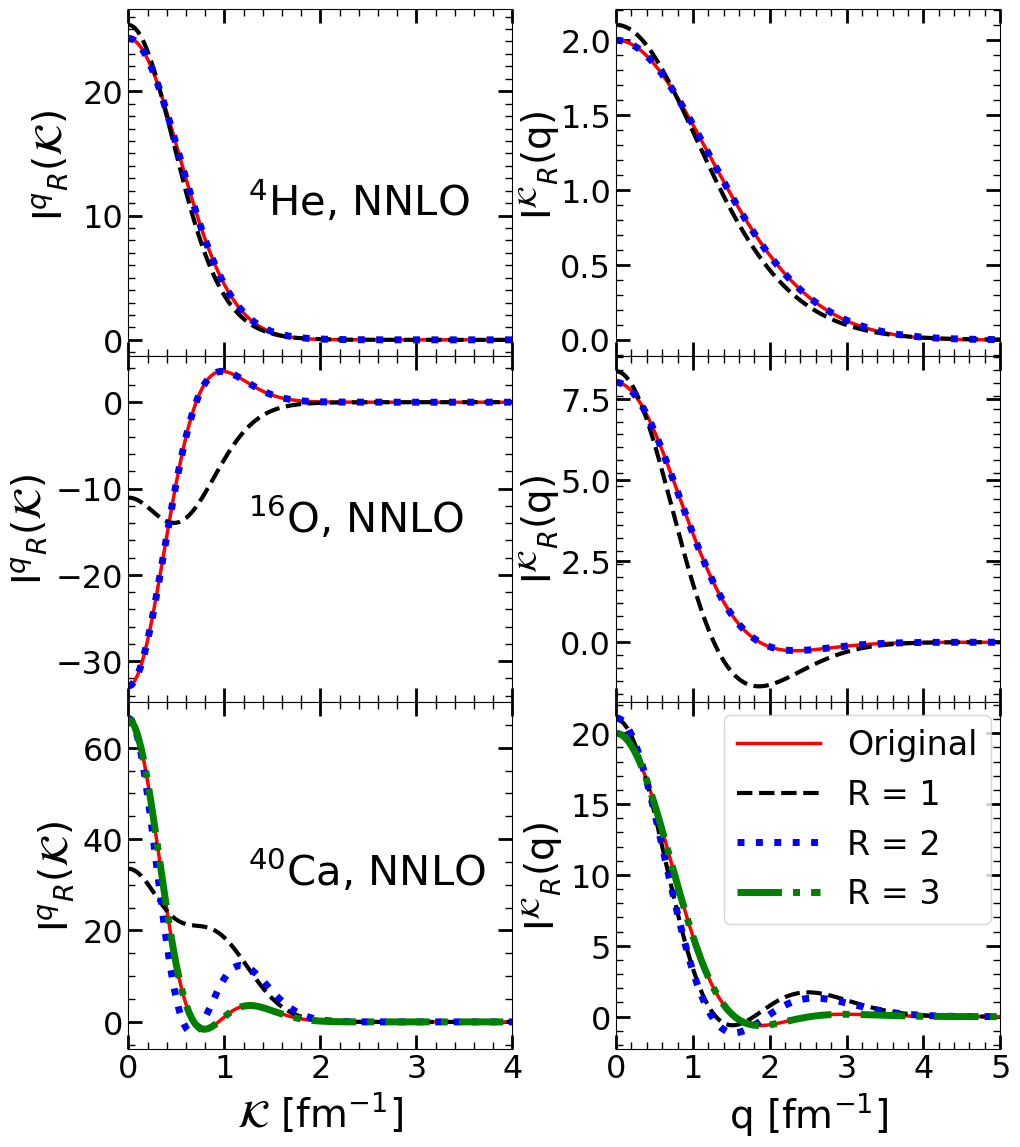}
\caption{The integrated neutron densities $I^{q}_{R}(\mathcal{K})$ (left column) and $I^{\mathcal K}_{R}(q)$ (right column) for the nuclei  $^{4}$He, $^{16}$O, $^{40}$Ca. The calculations are based on the NNLO$_{\rm opt}$ chiral potential and calculated from the off-shell densities shown in Fig.~\ref{fig4-1}. 
}
\label{fig4-2}
\end{figure}

For a more quantitative view on the separable approximations, we now turn to two integrated quantities that we can calculate, namely the integral of the off-shell density over $\mathcal{K}$, which gives us a function of $q$ that is, within a normalization factor, equivalent to the elastic form factor,
\begin{equation}
 I_R^{\mathcal{K}}(q) \equiv \int d\mathcal{K}\ \mathcal{K}^{2} \; \rho_R(q,\mathcal{K}) \,,
\label{eq:4.2}
\end{equation}
and the integral of the off-shell density over $q$, which gives us a function of $\mathcal{K}$
\begin{equation}
     I_R^q(\mathcal{K}) \equiv \int dq\ q^2 \; \rho_R(q,\mathcal{K}) \,.
\label{eq:4.3}
\end{equation}
These two dimensionless functions, calculated from the off-shell densities given in Fig.~\ref{fig4-1}, are shown in Fig.~\ref{fig4-2}; for the calculations based on the original off-shell densities $\rho(q,\mathcal{K})$, we omit the subscript $R$ which indicates the rank of the separable SVD approximation. 

Let us first concentrate on $^4$He, shown in the top panels of Fig.~\ref{fig4-2}.
Looking at $I_R^q(\mathcal{K})$ (left) and $I_R^{\mathcal{K}}(q)$ (right) shows that the rank-1 approximation is insufficient.  
This can be seen most easily at $I_R^{\mathcal{K}}(q=0)$, which should be exactly two for $^4$He, that is, the total number of neutrons (or protons, if we are considering the proton densities).
The minor correction coming from the second rank is necessary to satisfy this constraint and to accurately capture both $I_R^q(\mathcal{K})$ and $I_R^{\mathcal{K}}(q)$.  
Note that for $^4$He, the correction coming from the rank-2 term is relatively small for both integrated density functions.
Furthermore, at $N_{\rm max}=0$, a rank-1 approximation (not shown) is already sufficient to give the correct normalization at $q=0$.

Next, we consider the integrals over the integrated densities of the closed-shell nuclei $^{16}$O and $^{40}$Ca, shown in the middle and bottom panels of Fig.~\ref{fig4-2}.
It is immediately clear that a rank-1 separable ansatz for $\rho(q, \mathcal{K})$ cannot accurately represent either $I_R^q(\mathcal{K})$ (left) or $I_R^{\mathcal{K}}(q)$ (right) -- the shapes are quite different.  Also, just as for $^4$He,  the rank-1 approximation overshoots the correct value for the total number of neutrons at $I_R^{\mathcal{K}}(q=0)$.
In case of $^{16}$O, the rank-2 contribution correctly reproduces the full calculations for both $I_R^q(\mathcal{K})$ and $I_R^{\mathcal{K}}(q)$.
However, this is not the case for $^{40}$Ca -- for $^{40}$Ca we need at least a rank-3 separable ansatz to correctly represent these integrated quantities, which should not be surprising, given the bottom row of panels in Fig.~\ref{fig4-1}.  
Note that for $^{40}$Ca the rank-2 approximation does not seem to be very different from the rank-1 approximation for $I_R^{\mathcal{K}}(q)$, but for $I_R^q(\mathcal{K})$ the rank-1, rank-2, and rank-3 approximations have all three quite different shapes.

Furthermore, we want to point to an observation already made in Ref.~\cite{Burrows:2018ggt}, where we speculated that if a nucleus is dominated by $s$-shell nucleons, the value of $I_R^q(\mathcal{K}=0)$ is positive, but when $p$-shell nucleons dominate, $I_R^q(\mathcal{K}=0)$ is negative.  We reproduce these findings here for $^4$He and $^{16}$O, and we see that once the $sd$-shell is completely filled, as is the case for $^{40}$Ca, $I_R^q(\mathcal{K}=0)$ is again positive.  This suggests that when the highest filled shell has positive parity, $I_R^q(\mathcal{K}=0)$ is positive, but when the highest filled shell has negative parity, it is negative.

\begin{figure}[tb]
\centering
\includegraphics[width=0.99\columnwidth]{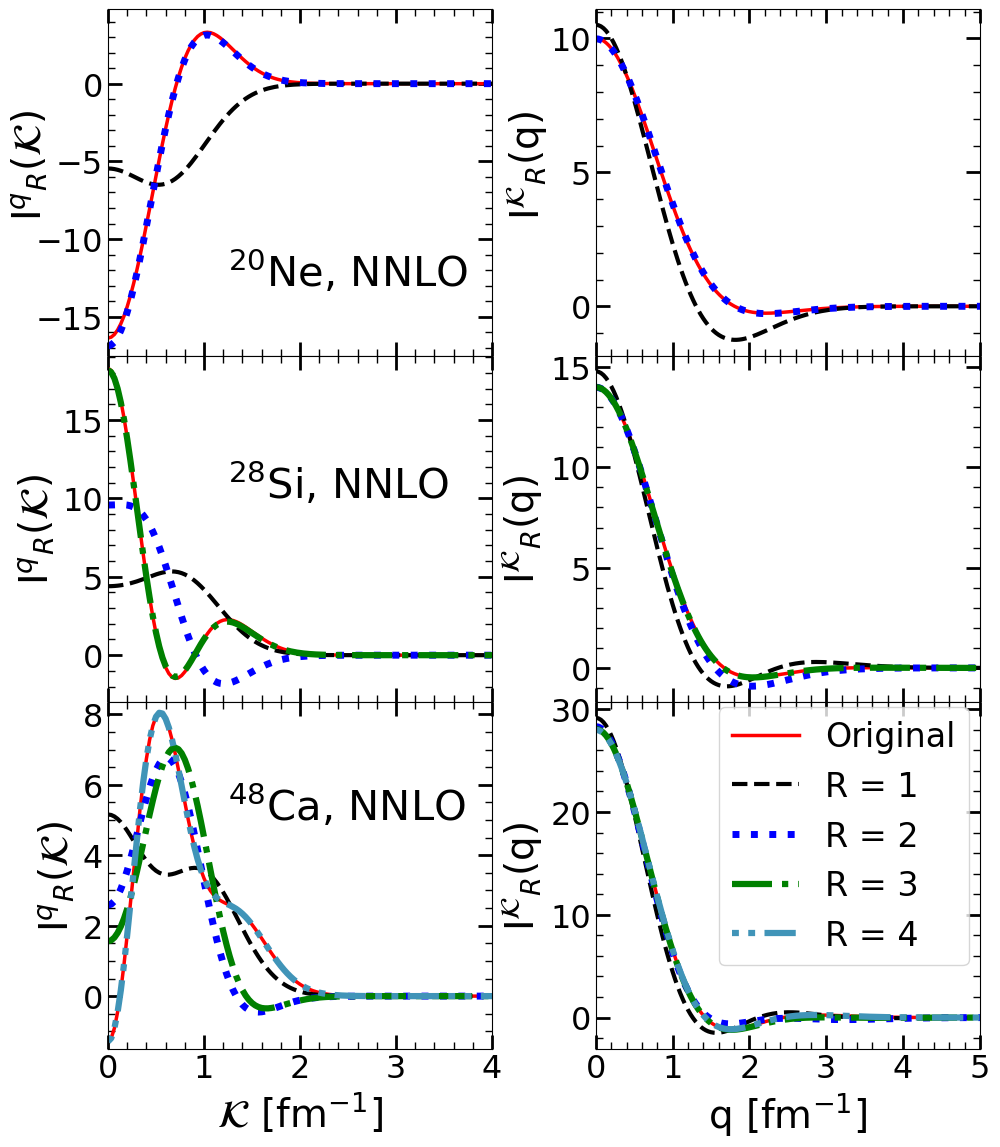}
\caption{The integrated neutron densities $I^q_{R}(\mathcal{K})$ (left) and $I^{\mathcal K}_{R}(q)$ (right) for the nuclei $^{20}$Ne ($N_{\rm max}=4$), $^{28}$Si ($N_{\rm max}=4$), and $^{48}$Ca ($N_{\rm max}=2$).  
The calculations are based on the NNLO$_{\rm opt}$ chiral potential. 
}
\label{fig4-3}
\end{figure}
Next, we show in Fig.~\ref{fig4-3} integrated densities $I_R^q({\mathcal K})$ and $I_R^\mathcal{K}(q)$ for the open-shell nuclei whose SVD eigenvalues are already shown in Fig.~\ref{svd-2}.
When looking at both $I_R^q({\mathcal K})$ and $I_R^\mathcal{K}(q)$ for $^{20}$Ne, we see that indeed a rank-2 approximation is sufficient to describe both of these integrated densities, which is consistent with the finding from Fig.~\ref{svd-2} in which $\eta(R=2)$ is below our chosen threshold for the tolerance. 
The normalization at $q=0$ does give for $^{20}$Ne $I_{R=2}^{\mathcal{K}}(q=0) = 9.99$, indicating that the SVD is a reliable tool to determine the rank of separability.
For $^{28}$Si, one can see that both $I_R^q({\mathcal K})$ and  $I_R^\mathcal{K}(q)$ need (at least) a rank-3 separable function to be described reasonably accurately.  This is most obvious when looking at the $I_R^q({\mathcal K})$. 
Finally, in the bottom row of Fig.~\ref{fig4-3} we see that for the 28 neutrons in $^{48}$Ca we need at least a rank-4 separable ansatz -- a rank-3 separable function cannot accurately describe the integrated density $I_R^q({\mathcal K})$. 
Note that we find the same behavior for both $I_R^q({\mathcal K})$ and $I_R^\mathcal{K}(q)$ when we use a chiral EFT interaction including $3N$ interactions, see Appendix \ref{appendixB}.

We also note that while $I_R^q({\mathcal K}=0)$ is negative for $^{20}$Ne, albeit less negative than for $^{16}$O (compare Fig.\ref{fig4-2}), for $^{28}$Si it becomes positive, though smaller than the corresponding value for $^{40}$Ca. 
This changes again when adding eight neutrons to fill the $f_{7/2}$-shell to obtain $^{48}$Ca, for which $I_R^q(\mathcal{K}=0)$ becomes slightly negative again. 
This seems to corroborate the conjecture from Ref.~\cite{Burrows:2018ggt} that the sign of $\rho(\mathcal{K}=0)$ may indicate whether even or odd parity shells are being filled.  More precisely, as a shell is being filled, the sign of $\rho(\mathcal{K}=0)$ changes from negative to positive if it is a positive parity shell, and the other way around for a negative parity shell.  However, it does not allow for assessing how many nucleons are occupying the valence shells.

A different quantitative measure of the accuracy of a separable approximation is how well it reproduces the calculated root-mean-square radius (RMS) of a given nucleus.  The RMS radius, $\sqrt{\langle \zeta^{2} \rangle}$, can be calculated from the local one-body density in coordinate space, $\tilde\rho(\zeta, Z=0)$, using
\begin{eqnarray}
\left \langle \zeta^{2} \right \rangle &=&  \frac{\sqrt{4\pi}}{N} \int \zeta^{4} \, \tilde{\rho}(\zeta, Z=0) \, d\zeta \,.
\label{eq:localRdensity}
\end{eqnarray}
\begin{table}[tb]
\let\mc\multicolumn
    \centering
\begin{tabular}
{|l|r||c|c|c|c|}
\hline
& $N_{\rm max}$ & Full & $R=1$ & $R=2$ & $R=3$ \\
\hline \hline 
$^{4}$He, NNLO$_{\rm opt}$  & 4 & 1.48 & 1.65 & 1.49 & 1.48\\
                            & 8 & 1.45 & 1.63 & 1.45 & 1.45\\
                            & 12 & 1.43 & 1.63 & 1.43 & 1.43\\
                            & 16 & 1.43 & 1.65 & 1.42 & 1.43\\
\hline  \hline 
$^{4}$He, Daejeon16     & 4 & 1.49 & 1.55 & 1.49 & 1.49\\
                        & 8 & 1.51 & 1.62 & 1.51 & 1.51\\
                        & 12 & 1.51 & 1.64 & 1.51 & 1.51\\
                        & 16 & 1.51 & 1.64 & 1.51 & 1.51\\
\hline  \hline 
$^{16}$O, NNLO$_{\rm opt}$ & 2 & 2.17 & 2.60 & 2.17 & 2.17\\
         & 6 & 2.20 & 2.69 & 2.20 & 2.20\\
         & 10 & 2.23 & 2.74 & 2.23 & 2.23\\
\hline  \hline 
$^{16}$O, Daejeon16 & 2 & 2.20 & 2.64 & 2.20 & 2.20\\
         & 6 & 2.27 & 2.72 & 2.28 & 2.27\\
         & 10 & 2.34 & 2.83 & 2.34 & 2.34\\
\hline  \hline 
$^{40}$Ca, NNLO$_{\rm opt}$ & 2 & 2.56 & 3.00 & 3.01 & 2.56\\
          & 4 & 2.58 & 3.00 & 3.01 & 2.58\\
          & 6 & 2.60 & 3.03 & 3.04 & 2.60\\
\hline  \hline 
$^{40}$Ca, Daejeon16 & 2 & 2.54 & 2.98 & 2.99 & 2.54\\
          & 4 & 2.56 & 2.97 & 2.98 & 2.56\\
          & 6 & 2.60 & 3.00 & 3.01 & 2.60\\ \hline  
\end{tabular}
\caption{The neutron RMS radii calculated from the exact densities $\rho_n(\zeta)$ and compared to their corresponding separable representations using ranks summed up to a given R as indicated in the table. All calculations use $\hbar\omega=20$~MeV.
}
\label{table-1}
\end{table}
The calculations for a fixed $\hbar\omega=20$~MeV for the three nuclei shown in Fig.~\ref{fig4-2} are listed in Table~\ref{table-1} as a function of $N_{\rm max}$ for two different interactions, namely the NNLO$_{\rm opt}$ and Daejeon16 interactions.
We indeed see that a rank-2 separable approximation can well represent $^4$He and $^{16}$O, while $^{40}$Ca needs a rank-3 representation, as measured by the accuracy of the corresponding RMS radius.  Again, the same conclusion holds when we include $3N$ interactions, see Appendix \ref{appendixB}.

\section{Singular vectors}
\label{eigenvectors}

In the previous Section, we studied how a separable approximation suggested by the SVD reproduces integrated densities as a function of the momentum transfer $q$ or the average momentum ${\mathcal K}$.  We verified that a separable approximation of rank $R$ represents the original densities quite well, with $R$ being the minimal value for which $\eta(R)$ falls below our threshold for the tolerance of $10^{-5}$, and where the actual value of $R$ increases with the number of nucleons.  We now want to go into more detail and study the leading singular vectors.
\begin{figure}[hbt]
\includegraphics[width=0.49\columnwidth]{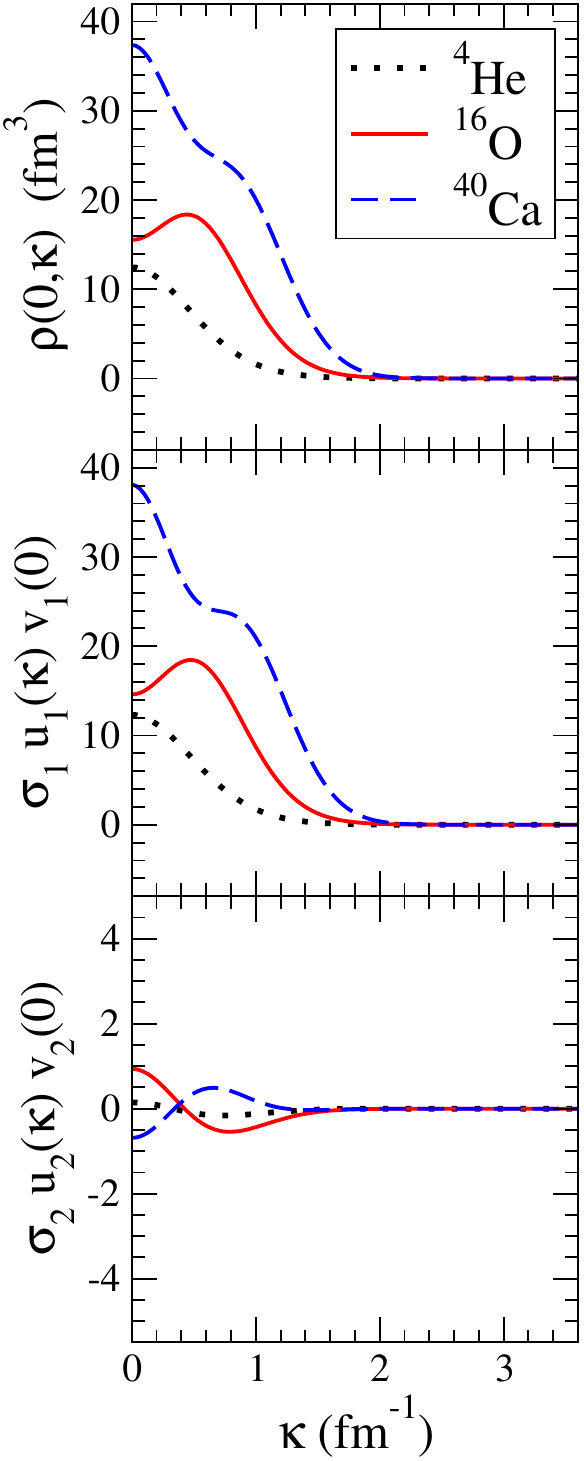}
\includegraphics[width=0.49\columnwidth]{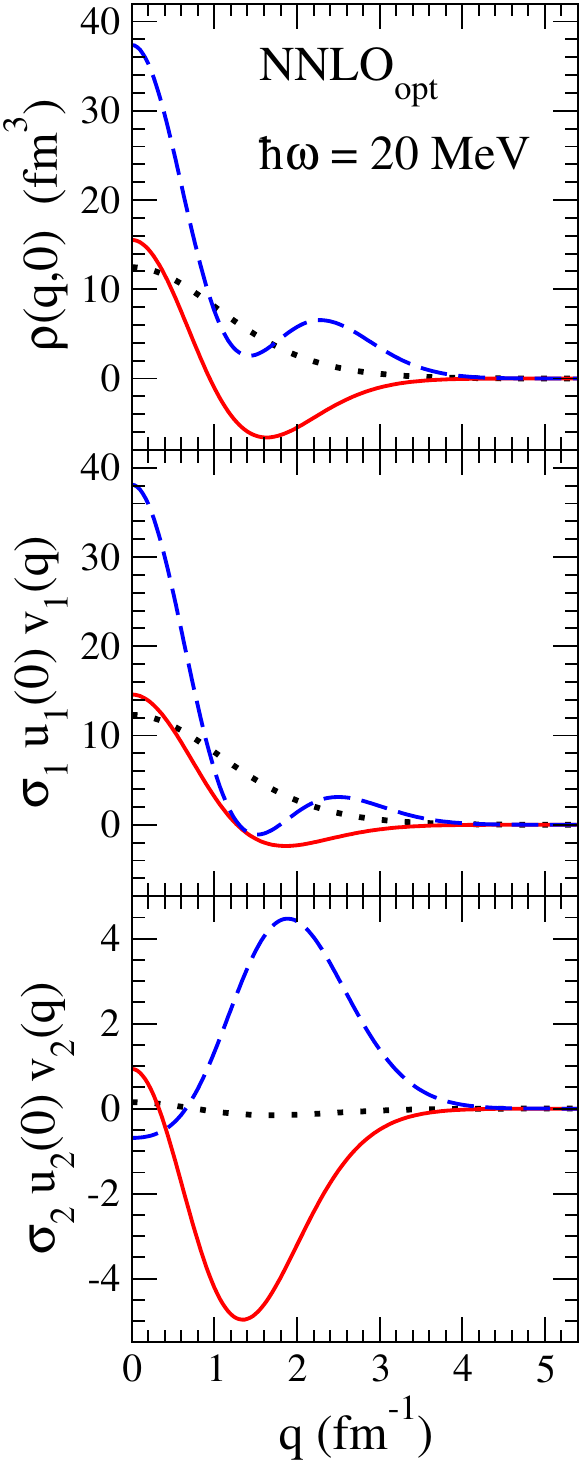}
\caption{The neutron densities $\rho(q=0, {\mathcal K})$ (left upper panel) and $\rho(q,{\mathcal K}=0)$ (right upper panel),
the left and right singular vectors ${\bf u}_1 ({\mathcal K})$ and ${\bf v}_1 (q)$, multiplied by the first singular value $\sigma_1$ as well as by ${\bf v}_1(0)$ and ${\bf u}_1(0)$ respectively (middle panels), and the left and right singular vectors ${\bf u}_2$ and ${\bf v}_2$, multiplied by the appropriate scaling factors (bottom panels).
\label{fig:rho_u1v1_u2v2}}
\end{figure}
In Fig.~\ref{fig:rho_u1v1_u2v2} we show in the top panels neutron one-body density distributions of $^4$He, $^{16}$O, and $^{40}$Ca, along the slices $\rho(0, \mathcal{K})$ (left) and $\rho(q, 0)$ (right), using the NNLO$_{\rm opt}$ interaction at $\hbar\omega=20$~MeV and the same values of $N_{\rm max}$ as in Fig.~\ref{svd-1} for these respective nuclei.  For ease of comparison, we use the same vertical scale for the left and right panels.
Before moving to the singular vectors, recall that in Sec.~\ref{nonlocald}, we already observed that the slice $\rho (0,{\mathcal K})$ is positive-definite, as can be seen from Fig.~\ref{fig2-2}, and that this slice can be identified with the probability of finding a neutron (or proton) with a momentum $p$ inside the nucleus.
In the top left panel of Fig.~\ref{fig:rho_u1v1_u2v2}, it is more clearly visible that this slice is indeed positive definite, as it should be in order to be interpreted as a probability distribution.
On the other hand, the slice $\rho(q, 0)$, shown in the top right panel, is not positive definite: for $^{16}$O it clearly drops below zero.
It is also interesting to note that the value at the origin, $\rho(0, 0)$, is nearly the same for $^4$He and $^{16}$O, but significantly larger for $^{40}$Ca.  We will come back to this in the next Section. 

In the middle and bottom panels of Fig.~\ref{fig:rho_u1v1_u2v2}, we show, on the left and right, respectively, the first two left singular vectors and the first two right singular vectors for the SVD of these closed shell nuclei $^4$He, $^{16}$O, and $^{40}$Ca. 
Since the left and right singular vectors themselves are normalized to one, we multiply them by the corresponding singular values, as well as by the value of the corresponding right and left singular vectors at the origin, in order to better judge their relative size.  
The middle panels clearly show that the leading singular vectors, or rather, $\sigma_1 {\bf u}_1(\mathcal{K}) {\bf v}_1(0)$ and $\sigma_1 {\bf u}_1(0) {\bf v}_1(q)$, have a similar behavior as the respective slices of the one-body density distributions $\rho(q, {\mathcal K})$ shown in the top panels.  
In particular $\sigma_1 {\bf u}_1(\mathcal{K}) {\bf v}_1(0)$ looks very similar to $\rho(0, \mathcal{K})$. We verified numerically that the ratio ${\bf u}_1(\mathcal{K}) / \rho(0,\mathcal{K})$ is approximately constant (i.e. independent of $\mathcal{K}$); only for calculations at $N_{\rm max}=0$ is it exactly constant.  
Thus, we may conclude that the first left singular vector of an SVD of the off-shell density can be interpreted as a good approximation to the momentum probability distribution for the corresponding nucleons in the nucleus.

On the other hand, although $\sigma_1 {\bf u}_1(0) {\bf v}_1(q)$ and $\rho(q, 0)$ look qualitatively similar, a more detailed inspection reveals that there are actually noticeable differences.  E.g, for $^{16}$O we see that $\rho(q, 0)$ drops well below zero for $q$ between one and two fm$^{-1}$, whereas $\sigma_1 {\bf u}_1(0) {\bf v}_1(q)$ falls much less below zero over this range.   And in contrast, for $^{40}$Ca the slice $\rho(q, 0)$ remains positive over this range, while $\sigma_1 {\bf u}_1(0) {\bf v}_1(q)$ drops slightly below zero for $q$ slightly larger than fm$^{-1}$.

The bottom panels of Fig.~\ref{fig:rho_u1v1_u2v2} show the second left and right singular vectors, multiplied by the second singular value and the value of the corresponding right and left singular vectors at the origin, that is, $\sigma_2 {\bf u}_2(\mathcal{K}) {\bf v}_2(0)$ and $\sigma_2 {\bf u}_2(0) {\bf v}_2(q)$.  
Clearly, they are less important than the leading singular vectors: these contributions (including appropriate scale factors) are significantly smaller than the leading contributions shown in the middle panels.  
In particular $\sigma_2 {\bf u}_2(\mathcal{K}) {\bf v}_2(0)$ in the lower left panel is very small, only a few percent of the leading term in the SVD expansion, $\sigma_1 {\bf u}_1(\mathcal{K}) {\bf v}_1(0)$ in the middle left panel; this is in perfect agreement with our observation above that $\sigma_1 {\bf u}_1(\mathcal{K}) {\bf v}_1(0)$ is already very similar to $\rho(0, \mathcal{K})$, and that the ratio of ${\bf u}_1(\mathcal{K}) / \rho(0,\mathcal{K})$ is nearly constant.

The contributions of $\sigma_2 {\bf u}_2(0) {\bf v}_2(q)$ however are of the order of 10\% compared the leading order term, $\sigma_2 {\bf u}_2(0) {\bf v}_2(q)$, as can be seen in the lower right panel of Fig.~\ref{fig:rho_u1v1_u2v2}.  
Here we can see that the second-order correction to the leading-order term for $^{16}$O is dominated by a large negative contribution between one and two fm$^{-1}$.  Adding this term to the leading-order term will bring the $R=2$ SVD approximation to $\rho(q, 0)$ much closer to $\rho(q, 0)$.  Similarly, the second-order correction to the leading-order term for $^{40}$Ca is dominated by a large positive contribution between one and three fm$^{-1}$; adding this correction to the leading-order term brings it closer to $\rho(q, 0)$.  

Finally, note that for $^4$He both the second-order term of $\sigma_2 {\bf u}_2(\mathcal{K}) {\bf v}_2(0)$ and that of $\sigma_2 {\bf u}_2(0) {\bf v}_2(q)$ are of the order of one percent or less than the leading order term. 
Indeed, even though $\eta(R=1)$ is larger than our tolerance threshold of $10^{-5}$, it is less than 0.1\%, so it is not surprising that the second-order contributions for $^4$He are so small.

\section{Considerations in an extreme shell model}
\label{extreme}

To further pursue our thoughts on the physical interpretation of the left singular value, $\sigma_1 {\bf u}_1$, we constructed nuclei in what we call the `extreme shell model', in which the individual sub-shells are successively filled.  In this model, the actual many-body wavefunctions can be represented by single Slater determinants.  
Note that we do not have an exact factorization of the intrinsic and the c.m. motion, since we are not applying an $N_{\max}$ truncation; however, we are only considering nuclei with at least $N=Z=20$ here, so any c.m. contribution (which scales with $1/A$) is ignored -- that is, we do not attempt to remove any approximate c.m. motion in this Section. 
The goal here is to investigate if the SVD approach also provides insight into these heavier systems, and whether or not any systematics found in the previous Sections for realistic NCSM calculations carry over into the realm of heavier nuclei that are currently not accessible to {\it ab initio} NCSM calculations. 

First, we use the single Slater determinants of select systems with closed sub-shells to construct the corresponding off-shell one-body densities, and carry out the SVD as in Sec.~\ref{svd}.  
\begin{figure}[tb]
\centering
\includegraphics[width=1.45\columnwidth]{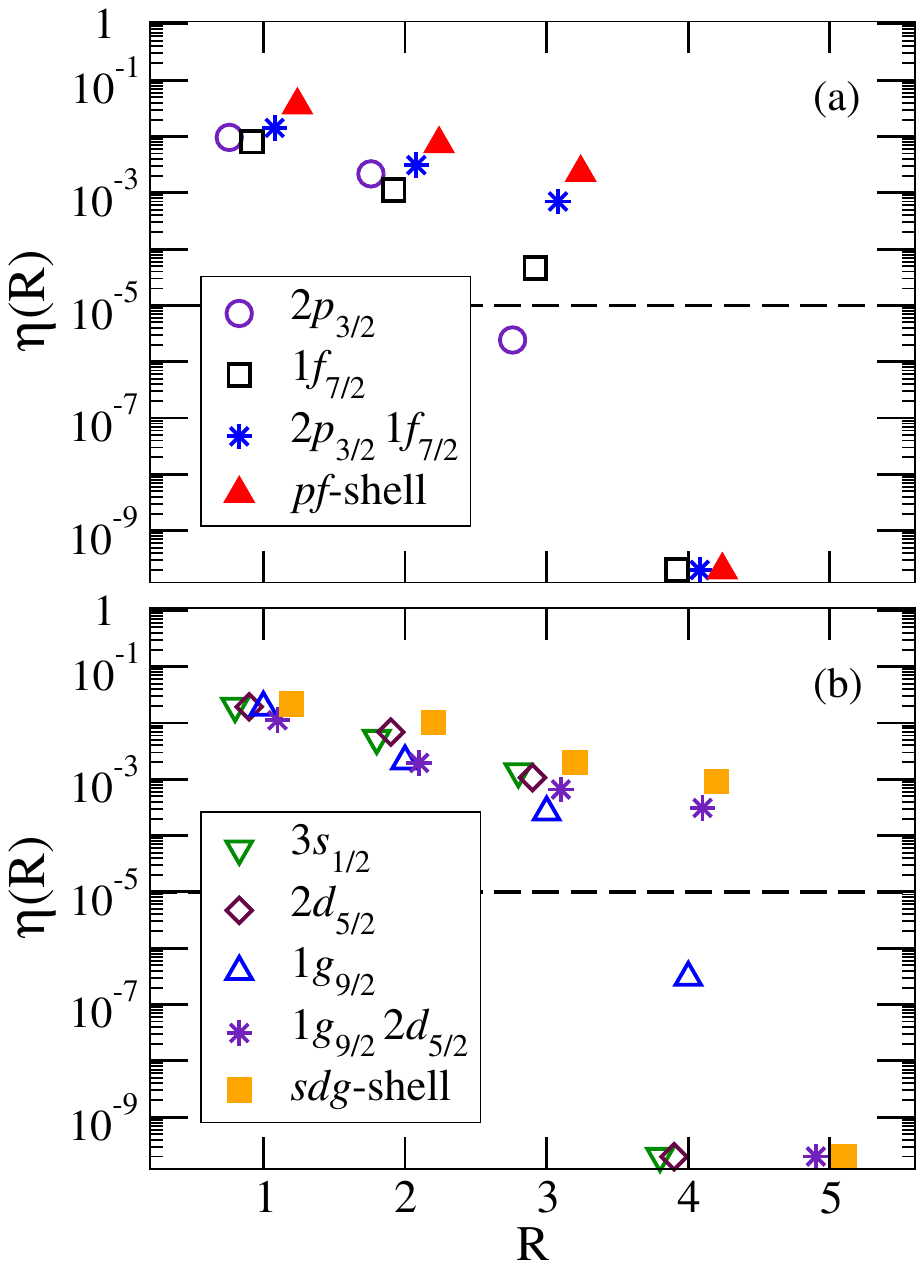}
\caption{The calculated values of $\eta(R)$ as a function of $R$ for calculations within an `extreme shell model', in which the (sub)shells are systematically filled, starting with the $2 p_{3/2}$ sub-shell (24 neutrons or protons) and ending with a filled $sdg$ shell (70 neutrons or protons).
}
\label{fig6-1}
\end{figure} 
In Fig.~\ref{fig6-1}, we start with nuclei that have a completely filled $sd$ shell (in addition to the lower shells, the $s$ and $p$ shells being filled). 
Using this as a core, we now fill separately the next-higher sub-shells, which are the $2p_{1/2}$, $2p_{3/2}$, $1f_{5/2}$, and $1f_{7/2}$ sub-shells. This allows us to investigate the effects of each sub-shell separately, or a combination of sub-shells, on the separability of the corresponding one-body density.
Adding four neutrons (or protons) into the $2p_{3/2}$ sub-shell (circles in the upper panel) gives $\eta(R)$ above the tolerance threshold when using only one or two singular values, but for $R=3$ the tolerance $\eta(R)$ drops just below our threshold of $10^{-5}$, indicating that a rank-3 separable representation of the one-body density is sufficient.  
On the other hand, with eight neutrons in the $1f_{7/2}$ sub-shell (squares in the upper panel) $\eta(R=3)$ remains just above our tolerance, and one would need a rank-4 separable representation.
Filling both the $1f_{7/2}$ and the $2p_{3/2}$ sub-shells, like for the neutrons in $^{58}$Fe, the stars in the upper panel indicate that for three singular values $\eta(R)$ is well above the tolerance threshold. In contrast, for $R=4$ its value falls below the threshold, indicating that a rank-4 separable function could represent nuclei with this configuration.  
Adding eight additional neutrons into the $1f_{5/2}$ and $2p_{1/2}$ does not lead to any further qualitative changes in the rank.  It may be illustrative to also consider the ratio of the extra nucleons in the $pf$ shell to the total number of nucleons one can have in the $pf$ shell, which is 20.
For the $2p_{3/2}$ sub-shell this ratio is 4:20, for the $1f_{7/2}$ sub-shell it is 8:20, where both ratios are smaller than 0.5, while for nucleons in both the $1f_{7/2}$ and the $2p_{3/2}$ sub-shells this ratio is 12:20, which is larger than 0.5. 

Moving to the lower panel of Fig.~\ref{fig6-1}, we define as core a fully filled $pf$ shell, and then add neutrons to the $3s_{1/2}$, $2d_{5/2}$, and $1g_{9/2}$ sub-shells, as well as the combined $2d_{5/2}$ and $1g_{9/2}$ sub-shells, and finally consider the complete $sdg$-shell. 
Here we see that either two neutrons in the $3s_{1/2}$ sub-shell,  or six in the $2d_{5/2}$ sub-shell, or ten in the $1g_{9/2}$ sub-shell gives $\eta(R)$ larger than the tolerance threshold for $R=1$, $2$, and $3$, but for $R=4$ the tolerance $\eta(R)$ drops below the threshold.  
This means that filling those sub-shells separately would require a rank-4 separable function to describe the one-body density.  
However, suppose we fill both the $2d_{5/2}$ and the $1g_{9/2}$ sub-shells with in total 16 neutrons. 
In that case, we see that $\eta(R=4)$ is still above the threshold of $10^{-5}$, and $\eta(R)$ drops below the threshold for $R=5$, indicating that for this case we need a rank-5 separable function to describe the one-body density. 
Additionally, a completely filled $sdg$ shell necessitates a rank-5 separable function to describe the one-body density accurately. 
In terms of the ratios of the number of neutrons over the total number of neutrons one can have in the $sdg$ shell, we see that as long as that ratio is below 0.5 (i.e., up to 15 nucleons -- the $sdg$ shell can have 30 neutrons) a rank-4 separable function should be sufficient. Once the shell is more than half-filled, one needs a rank-5 separable function to describe the one-body density accurately.

\begin{figure*}[tb]
\centering
\includegraphics[width=0.9\textwidth]{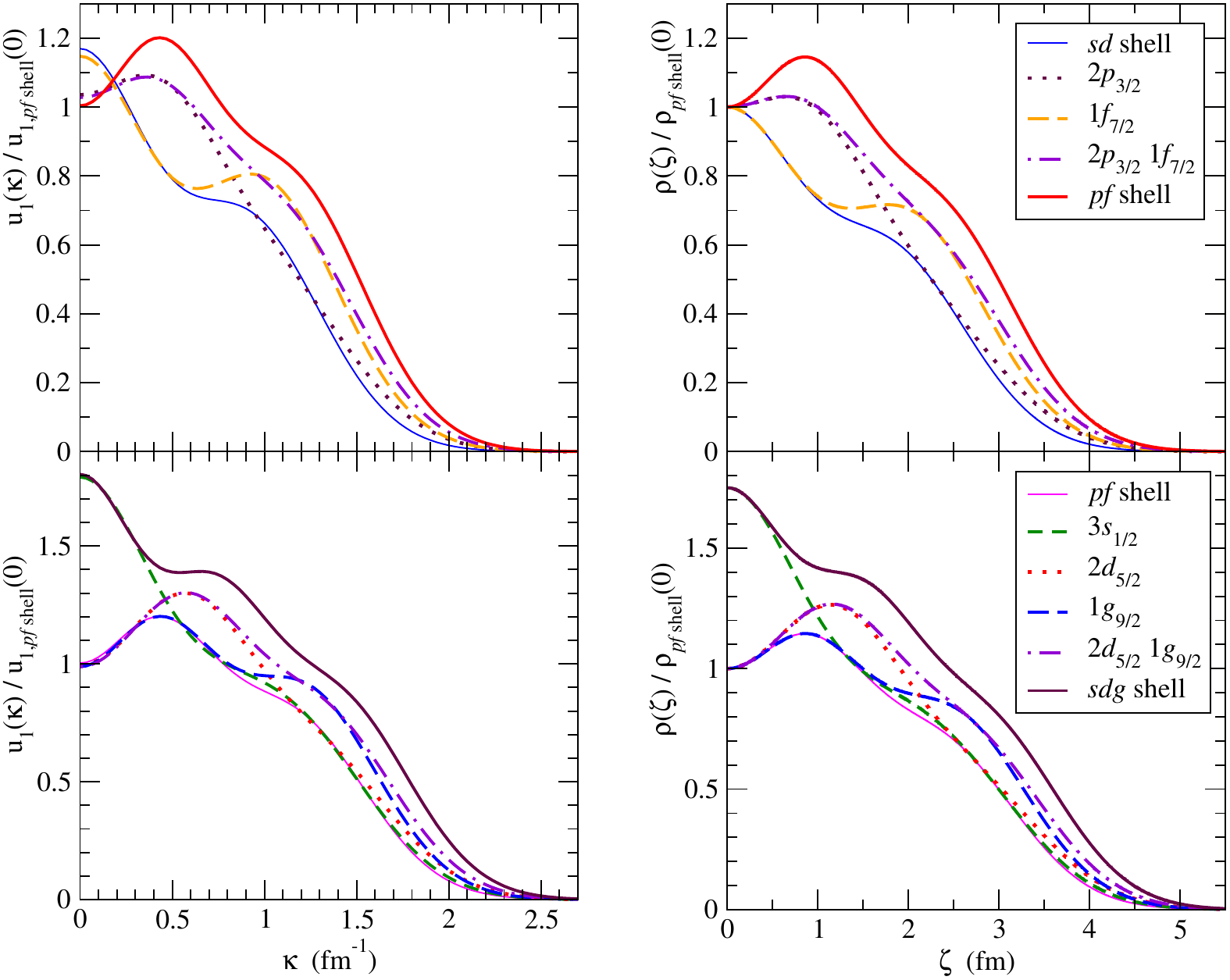}
\caption{The first left singular vectors as a function of ${\mathcal K}$ based on off-shell one-body densities obtained within the `extreme shell model',
corresponding to the SVD values shown in Fig.~\ref{fig6-1} (left), compared to the local one-body densities in coordinate space (right), both rescaled by their values at the origin for a closed $pf$ shell.  Top panels are results obtained for closed $sd$ shell, plus filled ${2}p_{3/2}$ (dotted), filled ${1}f_{7/2}$ (dashed), both filled ${2}p_{3/2}$ and filled ${1}f_{7/2}$ sub-shells, and closed $pf$ shell; bottom panels are for closed $pf$ shell, plus filled ${3}s_{1/2}$ (short-dashed), filled ${2}d_{5/2}$ (dotted), filled ${1}g_{9/2}$ (long dashed), both filled filled ${2}d_{5/2}$ and filled ${1}g_{9/2}$ sub-shells, and $sdg$ shell.}
\label{fig6-2}
\end{figure*} 
In Fig.~\ref{fig6-2} we show in the left two panels our results for the first left singular vectors $u_1({\mathcal K})$, computed in the 'extreme shell model',  corresponding to the SVD values that are shown in Fig.~\ref{fig6-2}, and in the right two panels we plot the corresponding local densities as function of $\zeta$ in coordinate space, $\tilde{\rho}(\zeta)$.  
In order to be able to compare the shapes, we divide the left singular vectors by $u_1(0)$ of the closed $pf$ shell, and similarly we divide $\tilde{\rho}(\zeta)$ by $\tilde{\rho}(0)$ of the closed $pf$ shell.  
The top two panels show our results for the $pf$ shell nuclei, whereas the bottom two panels contain the results for $sdg$ shell nuclei.
Note that since this is purely based on closed (sub-)shells in a HO basis, the shape of the local density distribution $\tilde{\rho}(\zeta)$ is the same as the shape of $\rho(0, \mathcal{K})$.

The first observation is how similar these leading singular vectors are to the corresponding local density distributions in coordinate space, even though one needs up to five singular vectors to accurately represent the corresponding off-shell density, to a tolerance of $10^{-5}$.  
This is in particular true for the $sdg$ shell (lower panels).
However, it is worth noting that already at $R=1$ the values of $\eta(R)$ are all between 0.01 and 0.04 for each of these cases, see Fig.~\ref{fig6-1} -- that is, with just the leading singular vectors, the preserved variance PV$(R=1)$ is already larger than 0.96.

Next, from these curves, it is evident that only the $s$-orbitals contribute at the origin, both for the singular vectors and for the local densities; all of the orbitals with nonzero angular momentum $L$ are zero at the origin.  More interesting is that the contributions from orbitals with different $L > 0$ are clearly distinguishable in both the leading singular vectors and in the local densities.  Again, not surprisingly, as $L$ increases, the peak contribution from the corresponding orbital shifts outwards.

\section{Conclusions}
\label{conclusions}

Motivated by recent findings on the separability of folding optical potentials for $NA$ elastic scattering, we study in this work the properties of off-shell nuclear one-body densities obtained in the NCSM framework. These one-body densities are one of the crucial ingredients for {\it ab initio} leading order optical potentials~\cite{Burrows:2018ggt,Burrows:2020qvu,Burrows:2023ygq,Baker:2023wla}. In this work, we concentrate on nuclei in the mass range $4\leq A \leq 48$ with $0^+$ ground states. Except for $^{48}$Ca, we concentrate on $N=Z$ nuclei. In a multipole expansion of the general one-body density, only the monopole term contributes to the nuclei as mentioned above. 

We start from the observation that this monopole term, when written in momentum variables {\bf q} and $\bm {\mathcal K}$, i.e., the momentum transfer and the average momentum, the off-shell density is independent of the angle between those two vectors. This is a good indication that $\rho^{(K=0)} ({\bf q}, {\bm {\mathcal K}})$ may be represented by separable functions in $q$ and $\mathcal K$. To determine if indeed the off-shell one-body density matrix can be represented by a low-rank separable approximation, we employ the SVD method. Using the singular values we define via a preserved variance a truncation error $\eta (R)$ to determine the minimum value $R$ for the rank for which a separable approximation can give an accurate representation of the off-shell one-body density of a given nuclear state; for convenience, we limit ourselves here to $0^+$ ground states. 

It turns out that the off-shell one-body density for nuclei with masses up to about $A=20$, or rather, $Z$ and/or $N=10$ can be approximated by rank-2 separable functions, whereas the densities for nuclei with $Z$ and/or $N$ between 12 and 24 are rank-3 separable.  Once either $Z$ or $N$ exceeds 24, a rank-4 separable approximation is required, as we illustrate for the neutron one-body density in $^{48}$Ca.
This statement does not depend on the size of the model space ($N_{\rm max}$) or the HO basis parameter ($\hbar \omega$) used in the NCSM, nor does it depend on the $NN$ interactions used in the calculation of the one-body density matrix.  This is even true when $3N$ interactions are included.  

To extend the mass range beyond the realm of the NCSM, we define an `extreme shell-model' in which sub-shells are successively filled beyond the $sd$-shell. In this simple approach, we find that a filled $pf$ shell may be approximated with rank-4 functions.  
Even filling an additional single $3s$-, $2d$-, or $1g$ sub-shell separately does not change this finding; only once the $sdg$-shell becomes approximately half-filled, we need to employ a rank-5 approximation for the one-body density.  
This simple study may suggest that even for high-mass nuclei, an off-shell one-body density matrix can be approximated with relatively low-rank separable functions, where the necessary rank can be estimated from the number of filled major shells. 
For the actual NCSM calculations, we explicitly show that the low-rank approximations are very accurate when calculating observables such as the RMS radius. 

We can also point to the physical interpretation of the $q$=0-slice of the off-shell one-body density matrix, $\rho (0,\mathcal K)$, as the probability of finding a nucleon with momentum ${\mathcal K}$.  It turns out that after the SVD, the first left singular vector is positive definite and within a scale factor approximately equal to the slice $\rho (0,\mathcal K)$.  A closer inspection of this first singular vector can give some information about the shell structure of the considered nucleus as we demonstrate for our `extreme shell-model' construction:  Not surprisingly, only $s$-shells contribute at ${\mathcal K} =0$ (in momentum representation) or at $\zeta=0$ (in coordinate representation), whereas for higher angular momentum sub-shells the peak in the probability distributions shifts to higher momenta or distances with increasing angular momentum.

In summary, the SVD is a highly effective tool for approximating off-shell density matrices of nuclei with $0^+$ ground states using low-rank, separable functions. We also verified that the monopole term of a nucleus with arbitrary spin in the ground state follows the same pattern as illustrated for the ground state of $^{10}$B in Appendix~\ref{appendixA}. These insights may help simplify numerical calculations, especially for higher mass nuclei.  Our finding of a universal separability of off-shell densities for nuclei with $0^+$ ground states corroborates the findings of Refs.~\cite{Arellano:2024xoc,Arellano:2022tsi} in the mass region for $A \geq 40$. The techniques applied here should lead to further investigations of {\it ab initio} optical potentials.


\begin{acknowledgments}
This work was performed in part under the auspices of the U.S.~Department of Energy under contract Nos.~DE-FG02-93ER40756 and DE-SC0023495 (SciDAC5/NUCLEI), and the National Science Foundation under Grant No. PHY-2310020. S.K.B. acknowledges the kind hospitality of the Institute of Nuclear and Particle Physics at Ohio University during his sabbatical visit. 
The authors acknowledge stimulating discussions with C. Drischler. 
The NCSM calculations for the results presented here were performed with the code MFDn \cite{doi:10.1002/cpe.3129,SHAO20181}.
The numerical computations benefited from computational resources provided by the National Energy Research Scientific Computing Center, a U.S.~Department of Energy Office of Science User Facility supported by the Office of Science of the U.S.~Department of Energy under Contract No. DE-AC02-05CH11231 using NERSC awards NP-ERCAP0028672, NP-ERCAP0032689, and NP-ERCAP0033451.
This research also used resources from the Argonne Leadership Computing Facility, a U.S. DOE Office of Science user facility at Argonne National Laboratory, which is supported by the Office of Science of the U.S. DOE under Contract No. DE-AC02-06CH11357, provided by the U.S. Department of Energy’s (DOE) Innovative and Novel Computational Impact on Theory and Experiment (INCITE) Program.

\end{acknowledgments}

\begin{appendix}
\section{SVD of monopole density of $^{10}$B}
\label{appendixA}
\begin{figure}
\centering
\includegraphics[width=0.99\columnwidth]{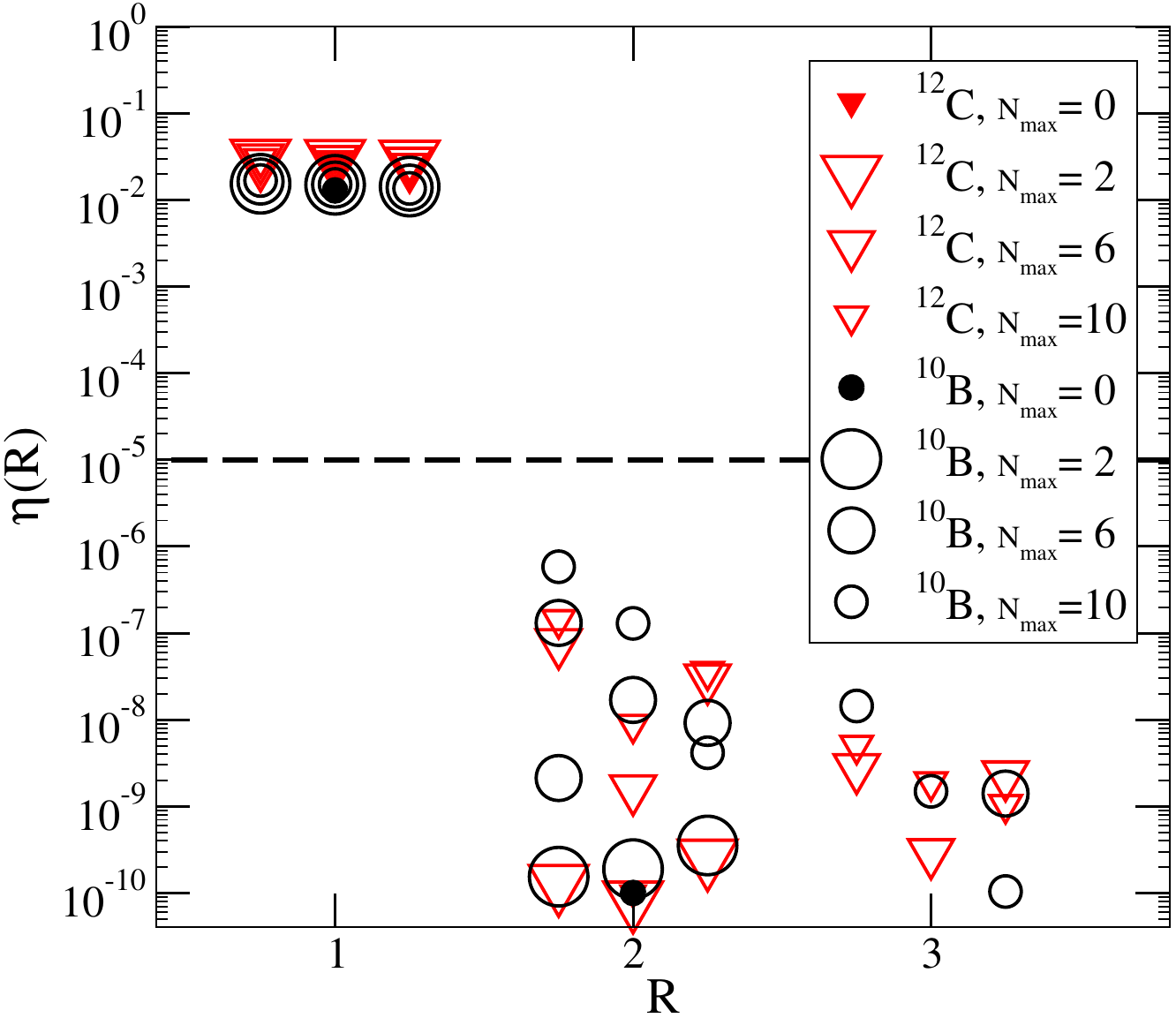}
\caption{The calculated value of $\eta(R)$ as function of $R$ for the monopole term of the $J^P = 3^+$ ground state of $^{10}$B, as well as the ground state of $^{12}$C, obtained with $N_{\rm max}= 0$, $2$, $6$, and $10$).  All calculations are based on the NNLO$_{\rm opt}$ chiral $NN$ potential.  The horizontal shifts of the symbols are explained in the text.
}
\label{fig:svd-b10} 
\end{figure}

In this appendix, we illustrate that for the $J^P=3^+$ ground state of $^{10}$B, the monopole term of the OBDM follows a similar separability pattern to that of $p$-shell nuclei with $0^+$ ground states, such as $^{12}$C.
In Fig.~\ref{fig:svd-b10} we show the SVD decompositions for the ground state neutron densities of $^{10}$B (circles) and $^{12}$C (triangles down) obtained from the NNLO$_{\rm opt}$ chiral $NN$ potential.  
The symbols are shifted horizontally as in Figs.~\ref{svd-1} and \ref{svd-2}:  Symbols centered on the major ticks represent an energy spacing of $\hbar\omega=20$~MeV,  whereas the symbols shifted to the left of the major tick represent $\hbar\omega=16$~MeV, and symbols shifted to the right represent $\hbar\omega=24$ MeV.

We see in Fig.~\ref{fig:svd-b10} that $\eta(R)$ behaves the same for these two densities: $\eta(R)$ is well above our threshold of $10^{-5}$ for $R=1$, but below this threshold for $R \ge 2$, indicating that these monopole densities are both rank-2 separable.  Furthermore, we see that the values of $\eta(R=1)$ are nearly independent of $N_{\max}$ and $\hbar\omega$ (starting from $N_{\max}=2$), whereas for $R \ge 2$ our results depend significantly on both $N_{\max}$ and $\hbar\omega$, while staying below the threshold of $10^{-5}$.  Indeed, it is natural that numerical noise increases with increasing basis size -- at $N_{\max}=10$ the size of the Hamiltonian matrix in the NCSM for these nuclei is several billion.  For $R \ge 4$ (not shown), our results for $\eta(R)$ for these two states are all well below $10^{-10}$.

\section{Effects of Three-Body Forces}
\label{appendixB}

In Sec.~\ref{svd} we showed that $3N$ interactions do not change the rank of separability of the nonlocal densities, see Fig.~\ref{svd-3}.  Here, we illustrate that the inclusion of $3N$ interaction does not qualitatively change the behavior of the separable representation either.  
\begin{figure}[t]
\centering
\includegraphics[width=0.99\columnwidth]{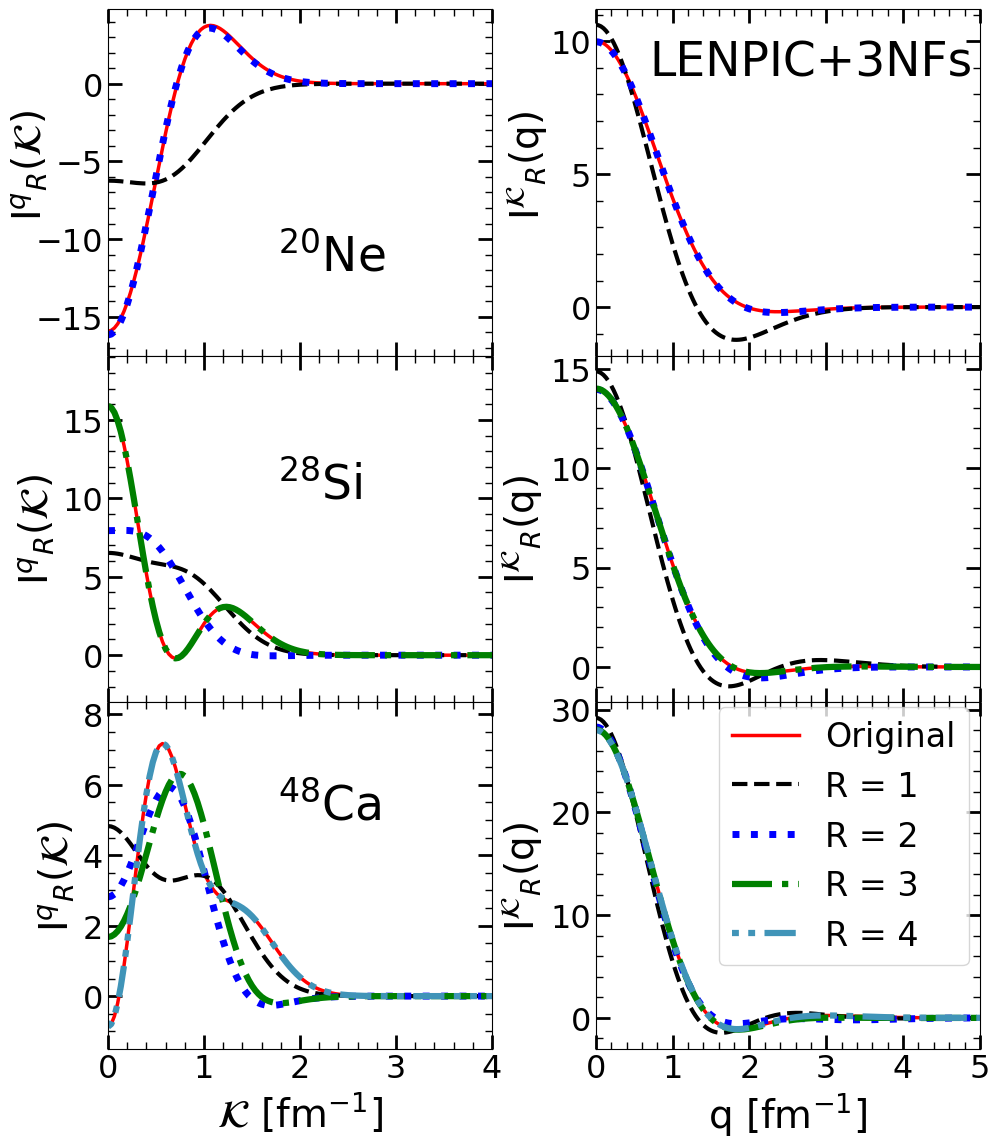}
\caption{The integrated neutron densities $I^q_{R}(\mathcal{K})$ (left) and $I^{\mathcal K}_{R}(q)$ (right) for the nuclei $^{20}$Ne ($N_{\rm max}=6$), $^{28}$Si ($N_{\rm max}=2$), and $^{48}$Ca ($N_{\rm max}=2$).  The calculations are based on the LENPIC SMS chiral EFT interaction at N$^3$LO, with a cutoff of $\Lambda = 500$~MeV, including $3N$ interactons \cite{Reinert:2017usi}.  The horizontal and vertical scales are the same as in Fig. \ref{fig4-3}. 
} 
\label{IntegratedQuants_LENPIC}
\end{figure}
In Fig.~\ref{IntegratedQuants_LENPIC} we show the integrated densities $I^{q}_{R}(\mathcal{K})$ and $I^{\mathcal K}_{R}(q)$ for the ground states of $^{20}$Ne, $^{28}$Si, and $^{48}$Ca, calculated with the LENPIC SMS interaction~\cite{Reinert:2017usi} with a chiral EFT cutoff of $\Lambda = 500$~MeV, SRG evolved to $\alpha = 0.08$~fm$^4$, including $3N$ interactions.  The six panels of Fig.~\ref{IntegratedQuants_LENPIC} look very similar to those of Fig. \ref{fig4-3}, which shows these same quantities obtained with the NNLO$_{\rm opt}$ chiral $NN$ potential,  demonstrating that the inclusion of $3N$ interactions does not change our conclusions regarding the rank of separable representations (note that the horizontal and vertical scales are the same as in Fig. \ref{fig4-3}).

\begin{table}[tb]
\let\mc\multicolumn
    \centering
\begin{tabular}
{|l|r||c|c|c|c|c|}
\hline
& $N_{\rm max}$ & Full & $R=1$ & $R=2$ & $R=3$ & $R=4$ \\
\hline \hline 
$^{16}$O    & 6 &  2.13  &  2.63  & 2.13  & 2.13  &  2.13 \\
$^{20}$Ne   & 6 &  2.26 &  2.78 &  2.25 &  2.26 &  2.26 \\
$^{28}$Si   & 2 &  2.36 &  2.79 &  2.34 &  2.36 &  2.36 \\
$^{40}$Ca   & 4 & 2.49  & 2.94  & 2.95  & 2.49  &  2.49 \\
$^{48}$Ca   & 2 &  2.62 &  2.92 &  2.71 &  2.63 &  2.62 \\
\hline
\end{tabular}
\caption{The neutron RMS radii calculated from the exact densities $\rho_n(\zeta)$ and compared to their corresponding separable representations using ranks summed up to a given R as indicated in the table. All calculations use $\hbar\omega=20$~MeV, and the LENPIC SMS interactions, including $3N$ interactions.
}
\label{table-appendixB}
\end{table}

Finally, in Table~\ref{table-appendixB} we show the neutron RMS radii for these three nuclei, as well as those of the closed shell nuclei $^{16}$O and $^{40}$Ca, all calculated with the LENPIC SMS interaction including $3N$ interactions.  Comparing this with Table~\ref{table-1}, we see that we need a rank-2 separable representation for $^{16}$O and a rank-3 representation for $^{40}$Ca, both with the $NN$-only interactions NNLO$_{\rm opt}$ and Daejeon16 (Table~\ref{table-1}), and with the $NN$ plus $3N$ LENPIC SMS interaction (Table~\ref{table-appendixB}).  Furthermore, for an accurate evaluation of the radii of the open $sd$-shell nuclei $^{20}$Ne and $^{28}$Si, we need a rank-3 representation. In contrast, for the neutron radius of $^{48}$Ca we need a rank-4 separable representation.
Looking at the values of $\eta(R)$ for these nuclei, see Fig.~\ref{svd-3}, we see that $\eta(R=3)$ is about $10^{-5}$ and $10^{-3}$ for $^{20}$Ne and $^{28}$Si, respectively, which makes it understandable why the difference in the radii at $R=2$ versus $R=3$ is only $0.4$\% and $1$\% for these two nuclei; and similarly, $\eta(R=4)$ is about $10^{-4}$ for $^{48}$Ca, with the change in the evaluated neutron radius being about $0.4$\% between $R=3$ and $R=4$ separable representations.   Note that for all three of these open-shell nuclei, the open shell is at most half-filled.

\end{appendix}

\clearpage


\bibliography{denspot,clusterpot,ncsm,reactions}



\end{document}